\pdfoutput=1
\RequirePackage{ifpdf}
\ifpdf 
\documentclass[pdftex]{sigma}
\else
\documentclass{sigma}
\fi

\DeclareMathOperator{\ad}{ad}
\DeclareMathOperator{\SC}{sc}
\DeclareMathOperator{\const}{const}
\DeclareMathOperator{\lwr}{lower}

\begin{document}

\newcommand{\arc}[1]{\mathrm{arc}(#1)}

\newcommand{\arXivNumber}{1407.5751}

\allowdisplaybreaks

\renewcommand{\PaperNumber}{020}

\FirstPageHeading

\ShortArticleName{Integrable Discrete Nonlinear Schr\"odinger Equation}

\ArticleName{Long-Time Asymptotics for the Defocusing Integrable Discrete Nonlinear Schr\"odinger Equation~II}

\Author{Hideshi YAMANE}

\AuthorNameForHeading{H.~Yamane}

\Address{Department of Mathematical Sciences, Kwansei Gakuin University,\\ Gakuen 2-1 Sanda, Hyogo 669-1337, Japan}

\Email{\href{mailto:yamane@kwansei.ac.jp}{yamane@kwansei.ac.jp}}

\URLaddress{\url{http://sci-tech.ksc.kwansei.ac.jp/~yamane/}}

\ArticleDates{Received September 06, 2014, in f\/inal form March 03, 2015; Published online March 08, 2015}

\Abstract{We investigate the long-time asymptotics for the defocusing integrable discrete nonlinear Schr\"odinger
equation.
If $|n|<2t$, we have decaying oscillation of order $O(t^{-1/2})$ as was proved in our previous paper.
Near $|n|=2t$, the behavior is decaying oscillation of order $O(t^{-1/3})$ and the coef\/f\/icient of the leading term is
expressed by the Painlev\'e II function.
In $|n|>2t$, the solution decays more rapidly than any negative power of~$n$.}


\Classification{35Q55; 35Q15}

\section{Introduction}

In our previous paper~\cite{IDNLS}, we studied the long-time behavior of the defocusing integrable discrete nonlinear
Schr\"odinger equation (IDNLS)
\begin{gather}
i \frac{d}{dt}R_n + (R_{n+1}-2R_n+R_{n-1})-|R_n|^2 (R_{n+1}+R_{n-1})=0
\label{eq:IDNLS}
\end{gather}
in the region $|n|\le (2-V_0)t$, $0<V_0<2$.
(In the present paper we refer to it as Region A.) We have proved that there exist $C_j=C_j(n/t)\in\mathbb{C}$ and
$p_j=p_j(n/t), q_j=q_j(n/t)\in\mathbb{R}$ ($j=1, 2$) depending only on the ratio $n/t$ such that
\begin{gather*}
R_n(t)=\sum\limits_{j=1}^2 C_j t^{-1/2} e^{- i (p_j t+q_j \log t)} +O\big(t^{-1}\log t\big)
\qquad
\text{as}
\quad
t\to\infty.
\end{gather*}
The behavior of each term in the sum is decaying oscillation of order $t^{-1/2}$.
Here $C_j$ and $q_j$ are def\/ined in terms of the ref\/lection coef\/f\/icient $r=r(z)$ (\cite{APT},~\cite{IDNLS})
corresponding to the initial potential $\{R_n(0)\}$.

In the present paper, we study~\eqref{eq:IDNLS} in other regions, namely one including the rays $n=\pm 2t$ and another
with $|n|>2t$.

Painlev\'e asymptotics has been observed in the cases of the MKdV equation (\cite{DZ}) and the Toda lattice
(\cite{Kamvissis}).
The proofs are based on the nonlinear steepest descent method.
Unlike the saddle point case, one has to deal with a~phase function of degree 3.
Following these results, especially~\cite{DZ}, we obtain the long-time asymptotics of~\eqref{eq:IDNLS} in Region B,
i.e.~near $n=\pm 2t$.

Roughly speaking, up to a~time shift $t \mapsto t-t_0$, our result is as follows (Theorem~\ref{th:Region0308}).

Consider a~curve def\/ined~by
\begin{gather}
t^{2/3} \frac{2-n/t}{(6-n/t)^{1/3}} =\textrm{a real constant}.
\label{eq:curve}
\end{gather}
It approaches $n/t=2$ with an error of $O(t^{-2/3})$ as $t\to\infty$.
The behavior of $R_n(t)$ on it is of the form
\begin{gather*}
R_n(t)=\const e^{i(-4t+\pi n)/2}t^{-1/3} +O\big(t^{-2/3}\big).
\end{gather*}
The constant in the above expression is written in terms of the Painlev\'e II function with parameters determined by the
ref\/lection coef\/f\/icient corresponding to $\{R_n(0)\}$.
A~similar result was obtained in~\cite{Nov} at least formally.
Notice that an analogous phenomenon can be found in a~dif\/ferent context~\cite{Kitaev}.
In the result of~\cite{DZ} about the MKdV equation, no oscillatory factor appears together with the Painlev\'e function.

\begin{remark}
\label{rem:lossofgenerality}
The equation~\eqref{eq:IDNLS} is invariant under the ref\/lection $n\mapsto -n$.
In the later sections, we assume $n>0$ without loss of generality.
\end{remark}

In Section~\ref{sec:Mainresults} we state our main results.
Sections~\ref{sec:Decomposition}--\ref{sec:Painleve} are devoted to the study of the region $2t-Mt^{1/3}<n<2t$.
In Section~\ref{sec:righthalf} we study $2t\le n<2t+M't^{1/3}$.
In Section~\ref{sec:RegionC} we investigate $n>2t$.

\section{Main results}
\label{sec:Mainresults}

Let $r(z)$ be the ref\/lection coef\/f\/icient determined by the initial potential $\{R_n(0)\}$.
See~\cite{IDNLS} for the precise def\/inition.
We assume that $\{R_n(0)\}$ decreases rapidly in the sense that for any $s>0$ there exists a~constant $C_s>0$ such that
$|R_n(0)|\le C_s/(1+|n|)^s$.\footnote{It is equivalent to saying that $\sum\limits_{s} (1+|n|)^s |R_n(0)|$ converges for
any~$s$, see~\cite{IDNLS}.}
Then $r(z)$ is smooth on $C\colon |z|=1$.

Let Region B\footnote{We only consider the case $n>0$, see Remark~\ref{rem:lossofgenerality} above.} be def\/ined~by
\begin{gather}
2t-M t^{1/3}< n< 2t+M' t^{1/3},
\label{eq:Region0308}
\end{gather}
where~$M$ and $M'$ are arbitrary positive constants.
The solution to an initial value problem for~\eqref{eq:IDNLS} has the following asymptotic behavior there:

\begin{theorem}
\label{th:Region0308}
Let $t_0$ be such that $ \pi^{-1}(\arg r(e^{-\pi i/4})-2t_0)-1/2$ is an integer.
Set $t'=t-t_0$, $p'=i(-4t'+\pi n)/4$, $\alpha'=[12t'/(6t'-n)]^{1/3}$, $q'=-2^{-4/3} 3^{1/3} (6t'-n)^{-1/3} (2t'-n)$.
Then we have
\begin{gather*}
R_n(t) =\frac{e^{2p'-\pi i/4} \alpha'}{(3t')^{1/3}} u\left(\frac{4q'}{3^{1/3}}\right) +O\big(t'^{-2/3}\big).
\end{gather*}
Here~$u$ is a~solution of the Painlev\'e II equation $u''(s)-su(s)-2u^3(s)=0$ and is specified
in~\eqref{eq:m8painleve}.
\end{theorem}

Let Region C be def\/ined~by
\begin{gather*}
n>(2+V_0)t,
\end{gather*}
where $V_0$ is an arbitrary positive constant.

\begin{theorem}
\label{th:Region0330}
Let~$j$ be an arbitrary positive integer.
Then in Region C, we have $R_n(t) =O(n^{-j})$.
More precisely, there exists a~constant $C=C(j, V_0)>0$ such that $|R_n(t)|\le C n^{-j}$ holds.

The solution decays exponentially if $r(z)$ is analytic on the circle $|z|=1$: there exists a~positive constant
$\rho=\rho(V_0)$ with $0<\rho<1$ and a~positive constant $C=C(V_0)$ such that $|R_n(t)|\le C \rho^n$ holds.
\end{theorem}

\begin{remark}
A~suf\/f\/icient condition for the analyticity of $r(z)$ is that $\{R_n(0)\}$ is f\/initely supported.
\end{remark}

\section{Decomposition and reduction}
\label{sec:Decomposition}
First we consider the long-time asymptotics in the `left-hand half' of Region B def\/ined in~\eqref{eq:Region0308}, namely
\begin{gather}
\label{eq:Region0308left}
2t-M t^{1/3}< n< 2t.
\end{gather}
Notice that a~curve like~\eqref{eq:curve} is in this kind of region for~$M$ suitably chosen.

Set
\begin{gather*}
\varphi =\varphi(z)=\varphi(z; n, t)=2^{-1} i t \big(z-z^{-1}\big)^2 - n\log z,
\qquad
\psi =\psi(z)=\varphi(z)/(it).
\end{gather*}
Choice of the branch of the logarithm is irrelevant because~$\varphi$ always appears in the form $e^{\pm\varphi}$.

We formulate a~Riemann--Hilbert problem (RHP):
\begin{gather}
m_+(z)=m_-(z) v(z)
\qquad
\text{on}
\quad
C\colon |z|=1,
\label{eq:originalRHP1}
\\
m(z)\to I
\qquad
\text{as}
\quad
z\to \infty,
\label{eq:originalRHP2}
\\
  v(z)=e^{-\varphi \ad\sigma_3}
\begin{bmatrix}
1-|r(z)|^2 & -\bar r(z)
\\
r(z) & 1
\end{bmatrix}.
\label{eq:originalRHP3}
\end{gather}
Here $m_+$ and $m_-$ are the boundary values from the \textit{outside} and \textit{inside} of~$C$ respectively of the
unknown matrix-valued analytic function $m(z)=m(z; n, t)$ in $|z|\ne 1$.
Namely~$C$ is endowed with clockwise orientation (a~convention adopted by~\cite{APT}).
We employ the usual notation $\sigma_3=\operatorname{diag}(1, -1)$, $a^{\operatorname{ad}\sigma_3}Q=a^{\sigma_3} Q a^{-\sigma_3}$
($a$: a~scalar,~$Q$: a~$2\times2$ matrix).
In formulating other RHPs in the remaining part of the present article, we will always assume the normalization
condition~\eqref{eq:originalRHP2}, which we often neglect to mention.

\begin{figure}[t]
\centering \includegraphics[width=5cm]{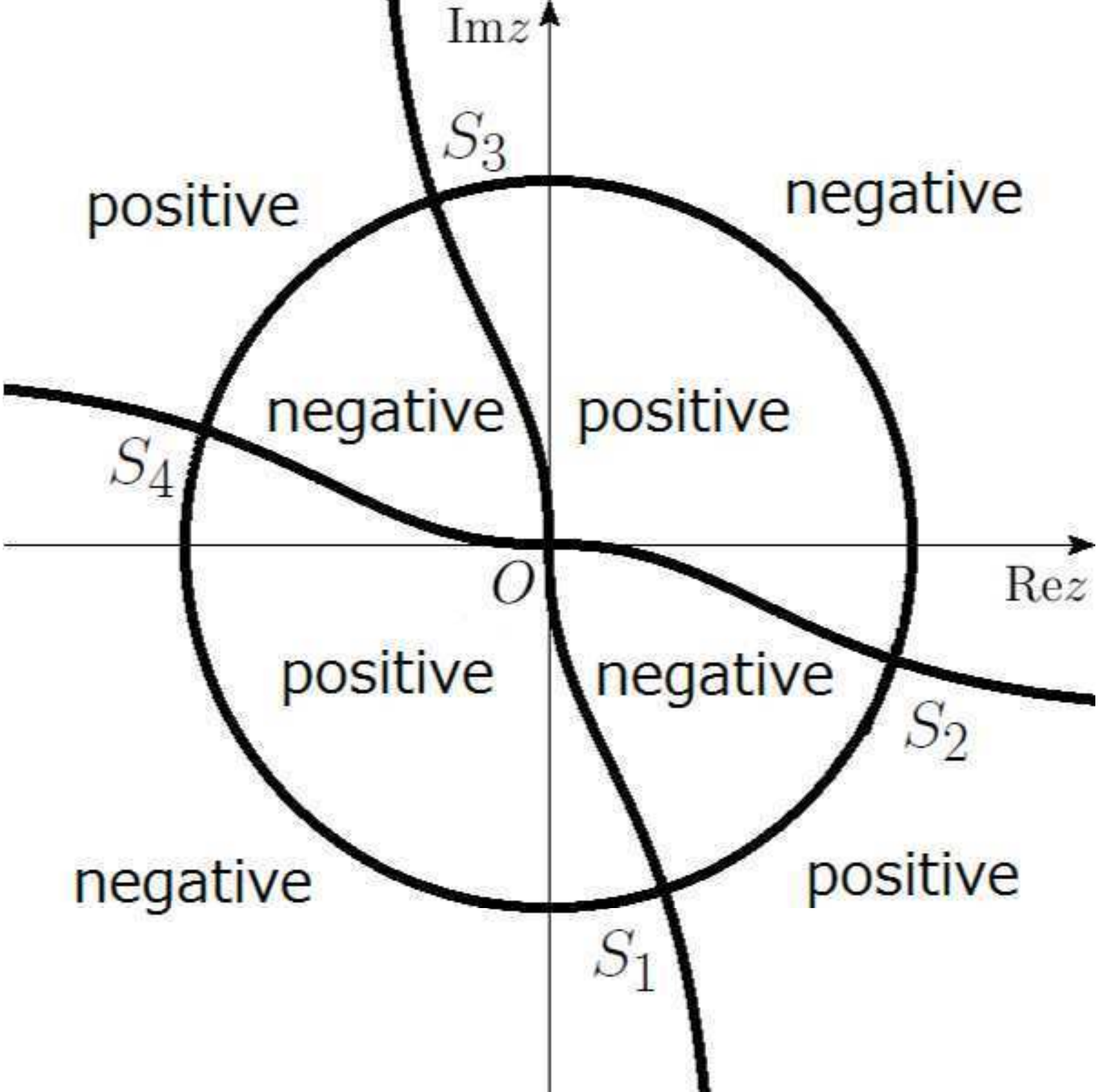}
 \caption{Real part of~$\varphi$.}
\label{fig:phase}
\end{figure}

The sign of the real part of~$\varphi$ is as in Fig.~\ref{fig:phase}.
The function $\varphi(z)$ has four saddle points.
They are
\begin{gather*}
S_1= e^{-\pi i/4}A,\qquad S_2= e^{-\pi i/4}\bar A,\qquad S_3=-S_1,\qquad S_4=-S_2,
\end{gather*}
where $A=2^{-1}(\sqrt{2+n/t }- i \sqrt{2-n/t})$.
One can reconstruct $\{R_n(t)\}$ from $m(z)$~by
\begin{gather}
R_n(t)=-\lim_{z\to 0} \frac{1}{z}m(z)_{21} =-\left.\frac{ d}{ d z} m(z)_{21} \right|_{z=0}.
\label{eq:Rnreconstruction}
\end{gather}
We will frequently use the factorization{\samepage
\begin{gather*}
v=v(z) =e^{-\varphi \ad \sigma_3} \left\{
\begin{bmatrix}
1 & -\bar r(z)
\\
0 & 1
\end{bmatrix}
\begin{bmatrix}
1 & 0
\\
r(z) & 1
\end{bmatrix}
\right\}
\end{gather*}
and its outcomes.}

For $\psi=\psi(z)=\varphi(z)/(it) $, we have
\begin{gather*}
\psi'(z)=z-z^{-3}-\frac{n}{it}z^{-1},
\qquad
\psi''(z)=1+3z^{-4}+\frac{n}{it}z^{-2},
\qquad
\psi'''(z)=-12z^{-5}-\frac{2n}{it}z^{-3}.
\end{gather*}
Third-order approximation of~$\psi$ will be necessary, since we will deal with coalescence of saddle points.

We do not need the `$\Delta$-conjugation' as in~\cite[\S~4]{IDNLS}, where a~function called~$\rho$ was introduced.
Here we decompose $\bar r$ and~$r$ on $\arc{S_2 S_3} \cup \arc{S_4 S_1}$ by using Taylor's theorem and Fourier
analysis\footnote{We sometimes denote $\arc{S_j S_k}$ by $S_j S_k$.}.

Set $\vartheta=\theta-\pi/4$, $z=e^{i\theta}$ and $\vartheta_0=\pi/2+\arg A=\pi/2-\arctan\sqrt{(2t-n)/(2t+n)}$.
(The def\/initions of~$\vartheta$ and $\vartheta_0$ are dif\/ferent from those in~\cite{IDNLS}.) Then $\arc{S_2 S_3}$
corresponds to $-\vartheta_0\le \vartheta\le \vartheta_0$.
We regard the function $\bar{r}$ on $\arc{S_2 S_3}$ as a~function in~$\vartheta$ and denote it by $\bar{r}(\vartheta)$
by abuse of notation.
We have
\begin{gather*}
\bar{r}(\vartheta)=H_{\rm e}\big(\vartheta^2\big)+\vartheta H_{\rm o} \big(\vartheta^2\big), \qquad -\vartheta_0\le \vartheta \le \vartheta_0,
\end{gather*}
for smooth functions $H_{\rm e}$ and $H_{\rm o}$.
By Taylor's theorem, they are expressed as follows:
\begin{gather*}
H_{\rm e}\big(\vartheta^2\big) =\mu_0^{\rm e}+\cdots+\mu_k^{\rm e}\big(\vartheta^2-\vartheta_0^2\big)^k +\frac{1}{k!} \int_{\vartheta_0^2}^{\vartheta^2}
H_{\rm e}^{(k+1)}(\gamma) \big(\vartheta^2-\gamma\big)^k d \gamma,
\\
H_{\rm o}\big(\vartheta^2\big) =\mu_0^{\rm o}+\cdots+\mu_k^{\rm o}\big(\vartheta^2-\vartheta_0^2\big)^k +\frac{1}{k!} \int_{\vartheta_0^2}^{\vartheta^2}
H_{\rm o}^{(k+1)}(\gamma) \big(\vartheta^2-\gamma\big)^k d \gamma.
\end{gather*}
Here $k=4q+1$ and~$q$ can be any positive integer.

We set
\begin{gather*}
R(\vartheta) =R_k(\vartheta) =\sum\limits_{i=0}^k \mu_i^{\rm e} \big(\vartheta^2-\vartheta_0^2\big)^i +\vartheta\sum\limits_{i=0}^k
\mu_i^{\rm o} \big(\vartheta^2-\vartheta_0^2\big)^i,
\\
\alpha(z) =(z-S_2)^q (z-S_3)^q,
\qquad
h(\vartheta) =\bar{r}(\vartheta)-R(\vartheta)
\end{gather*}
and, by abuse of notation,
\begin{gather*}
\alpha(\vartheta) =\alpha\big(e^{i (\vartheta+\pi/4)}\big)
=\big[e^{i (\vartheta+\pi/4)}- e^{i (-\vartheta_0+\pi/4)}\big]^q \big[e^{i(\vartheta+\pi/4)}- e^{i (\vartheta_0+\pi/4)}\big]^q.
\end{gather*}
Notice that we have $R(\pm \vartheta_0)=\bar{r}(\pm \vartheta_0) $.
The function~$R$ extends analytically from $\arc{S_2S_3}$ to a~complex neighborhood.
By abuse of notation, $R(z)$ denotes the analytic function thus obtained, so that
$R(e^{i(\vartheta+\pi/4)})=R(\vartheta)$ and $R(S_j)=\bar{r}(S_j)$.

On $\arc{S_2 S_3}$ we have $d\psi/d\vartheta=-2\cos2\vartheta-n/t$.
Since $[-\vartheta_0, \vartheta_0]\ni\vartheta\mapsto \psi\in\mathbb{R}$ is strictly decreasing, we can consider its
inverse $\vartheta=\vartheta(\psi)$, $\psi(\vartheta_0)\le\psi\le\psi(-\vartheta_0)$.
We set
\begin{gather*}
(h/\alpha)(\psi)=
\begin{cases}
h(\vartheta(\psi))/\alpha(\vartheta(\psi)) & \text{if}
\ \
\psi(\vartheta_0)\le \psi \le \psi(-\vartheta_0),
\\
0 & \text{otherwise}.
\end{cases}
\end{gather*}

\looseness=1
Then $(h/\alpha)(\psi)$ is well-def\/ined for $\psi\in\mathbb{R}$.
It can be shown that $h/\alpha\in H^p (-\infty<\psi<\infty)$, where~$p$ can be any positive integer if we choose
a~suf\/f\/iciently large value of~$k$.
Its norm is uniformly bounded with respect to $(n, t)$.
This argument is a~`curved' version of~\cite[equation~(1.33)]{DZ}.

Notice that $d\psi/d\vartheta=-2\cos2\vartheta-n/t$ has a~zero of order \textit{two} at $\vartheta=\pm\pi/2$ if $n/t=2$.
It may worsen the estimate of the Sobolev norm (cf.~\cite[equation~(1.33)]{DZ}) of $h/\alpha$ as a~function of~$\psi$ in contract
to~$\vartheta$, especially in the case of Section~\ref{sec:righthalf}, since it involves
$d/d\psi=(d\psi/d\vartheta)^{-1} d/d\vartheta$.
This kind of dif\/f\/iculty is overcome by choosing a~suf\/f\/iciently large value of~$k$.

Set
\begin{gather*}
 (\widehat{h/\alpha})(s) = \int_{-\infty}^{\infty} e^{- i s \psi} (h/\alpha)(\psi) \frac{ d \psi}{\sqrt{2\pi}},
\\
 h_{\rm I} (\vartheta) =\alpha(\vartheta) \int_t^\infty e^{i s \psi(\vartheta)} (\widehat{h/\alpha})(s) \frac{ d
s}{\sqrt{2\pi}},
\\
 h_{\rm II} (\vartheta) =\alpha(\vartheta) \int_{-\infty}^t e^{i s \psi(\vartheta)} (\widehat{h/\alpha})(s) \frac{ d
s}{\sqrt{2\pi}},
\end{gather*}
then we have $h(\vartheta)=h_{\rm I}(\vartheta)+h_{\rm II}(\vartheta)$ and
$\bar{r}(\vartheta)=R(\vartheta)+h_{\rm I}(\vartheta)+h_{\rm II}(\vartheta)$ on $|\vartheta|\le\vartheta_0$.
We can apply the same process to~$r$.
We have
\begin{alignat}{3}
& \bar r(z) =\bar r=h_{\rm I}+h_{\rm II}+R,
\qquad&&
r(z) =r=\bar h_{\rm I}+\bar h_{\rm II}+\bar R,&\nonumber
\\
&\bar r(S_j)=R(S_j),
\qquad&&
r(S_j)=\bar R(S_j).&
\label{eq:decomp}
\end{alignat}

The decomposition on $\arc{S_4 S_1}$ immediately follows by symmetry.
Notice that $h_{\rm II}$, $R$, $\bar h_{\rm II}$ and $\bar R$ can be analytically continued to certain open sets.
We still employ the same notation for the extended functions.
For example, $\bar h_{\rm II}=\bar h_{\rm II}(z)$ is analytic, although the bar may seem a~little strange.

\begin{figure}[t]\centering
\begin{minipage}[b]{61mm}
\centering \includegraphics[width=6cm]{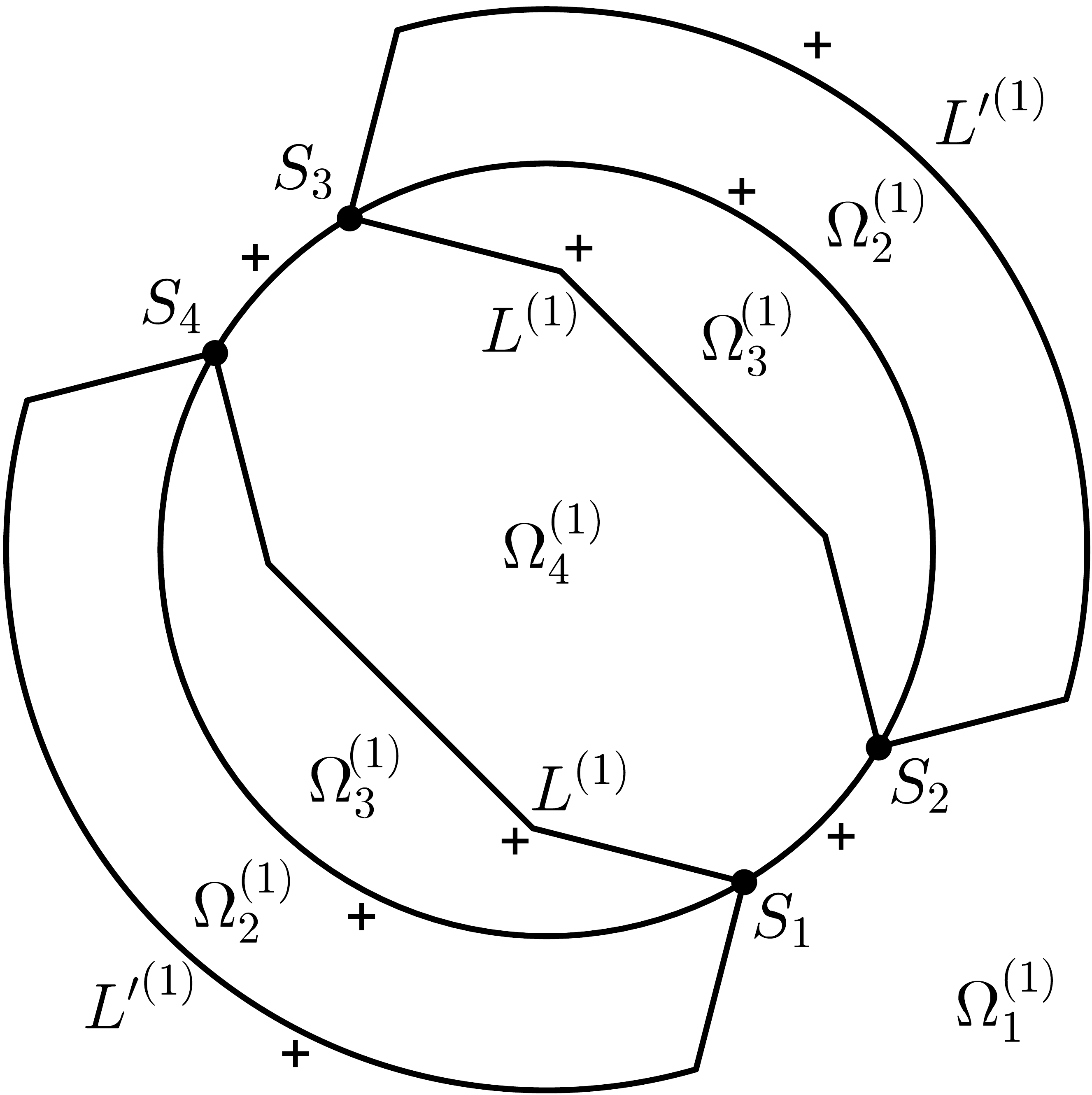} \caption{$\Sigma^{(1)}$.}
\label{fig:Sigma1}
\end{minipage}\qquad\qquad
\begin{minipage}[b]{33mm}
\centering \includegraphics[width=3.2cm]{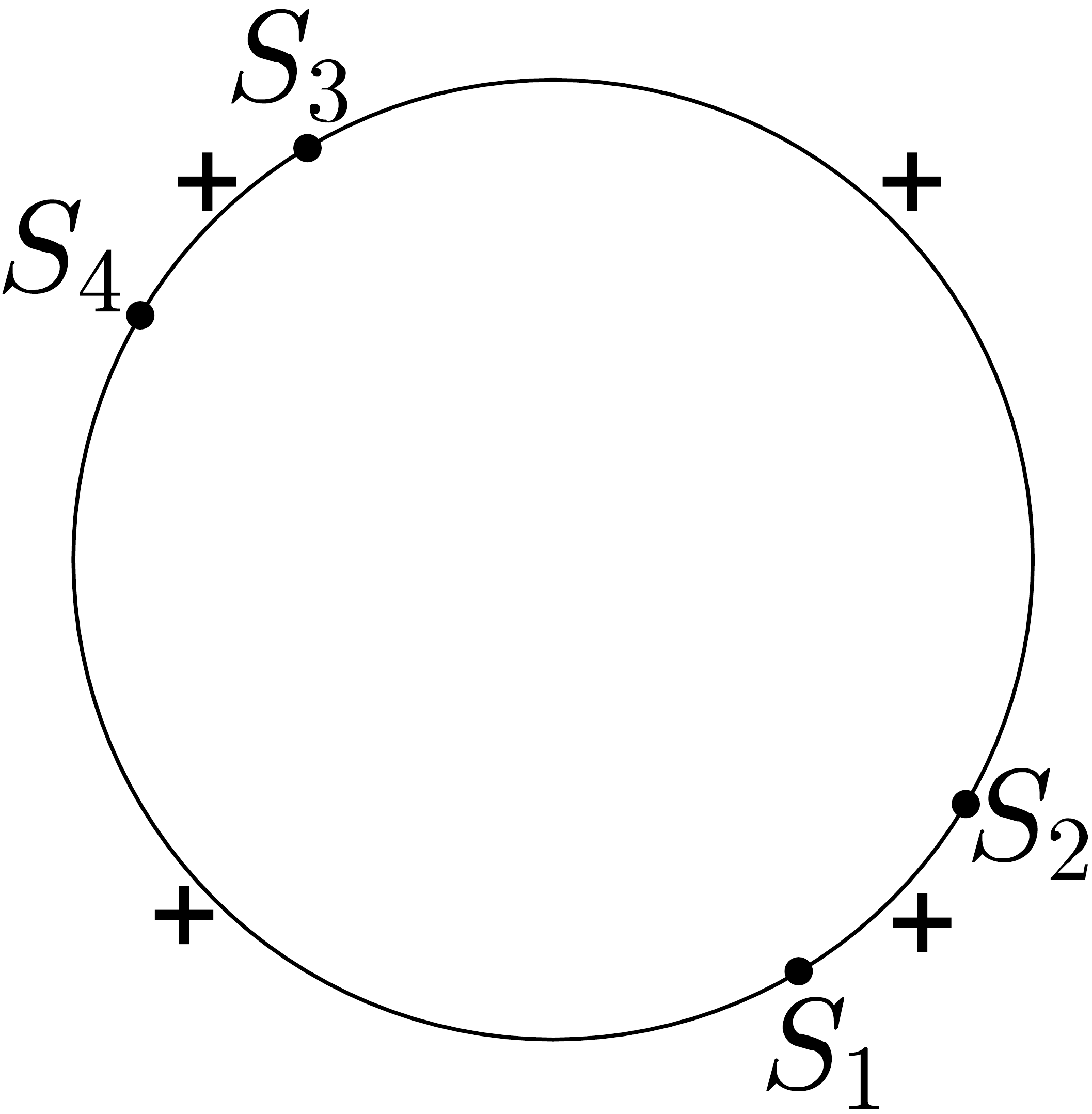} \caption{$\Sigma^{(2)}$.}
\label{fig:Sigma2}
\end{minipage}
\end{figure}

We introduce a~new contour $\Sigma^{(1)}$ as in Fig.~\ref{fig:Sigma1}.
It is a~variation of~$\Sigma$ in~\cite[Fig.~2]{IDNLS}.
The part~$L^{(1)}$ is bent so that it stays away from the circle as $n\to 2t$ (except near the saddle points.) Some open
sets, not necessarily connected, are def\/ined in Fig.~\ref{fig:Sigma1}.
Notice that~$\Sigma^{(1)}$ remains f\/inite even as $n\to 2t$.

We introduce a~new unknown matrix $m^{(1)}$ by setting
\begin{gather*}
m^{(1)}= \begin{cases} m
& \text{in}
\ \
\Omega_1^{(1)}\cup \Omega_4^{(1)},
\\
  m e^{-\varphi \ad \sigma_3}
\begin{bmatrix}
1 & 0
\\
-\bar h_{\rm II} & 1
\end{bmatrix}
&
\text{in}
\ \
\Omega_2^{(1)},
\vspace{1mm}\\
 m e^{-\varphi \ad \sigma_3}
\begin{bmatrix}
1 & -h_{\rm II}
\\
0 & 1
\end{bmatrix}
&
\text{in}
\ \
\Omega_3^{(1)}.
\end{cases}
\end{gather*}
We def\/ine a~new jump matrix $v^{(1)}$~by
\begin{gather*}
v^{(1)}=v= \begin{cases}
 e^{-\varphi \ad \sigma_3} \left\{
\begin{bmatrix}
1&-\bar r
\\
0 & 1
\end{bmatrix}
\begin{bmatrix}
1 & 0
\\
r & 1
\end{bmatrix}
\right\}
&
\text{on}
\ \
S_1S_2\cup S_3S_4,
\vspace{1mm}\\
 e^{-\varphi \ad \sigma_3}
\begin{bmatrix}
1 & -h_{\rm II}
\\
0 & 1
\end{bmatrix}
&
\text{on}
\ \
L^{(1)},
\vspace{1mm}\\
  e^{-\varphi \ad \sigma_3} \left\{
\begin{bmatrix}
1 & -h_{\rm I} - R
\\
0 & 1
\end{bmatrix}
\begin{bmatrix}
1 & 0
\\
\bar h_{\rm I} + \bar R & 1
\end{bmatrix}
\right\}
&
\text{on}
\ \
S_2S_3\cup S_4S_1,
\vspace{1mm}\\
 e^{-\varphi \ad \sigma_3}
\begin{bmatrix}
1 & 0
\\
\bar h_{\rm II} & 1
\end{bmatrix}
&
\text{on}
\ \
L'^{(1)}.
\end{cases}
\end{gather*}
Then we have
\begin{gather*}
  m^{(1)}_+=m^{(1)}_- v^{(1)}
\qquad
\text{on}
\quad
\Sigma^{(1)},\qquad
  m^{(1)} \to I
\qquad
\text{as}
\quad
z\to\infty.
\end{gather*}
By~\eqref{eq:Rnreconstruction}, we have
\begin{gather}
\label{eq:reconstruction}
R_n (t)=-\frac{d}{dz} m^{(1)}(z)_{21}\bigg|_{z=0}.
\end{gather}

We need some estimates.
First, in the same way as~\cite[equation~(1.36)]{DZ} and~\cite[equation~(43)]{IDNLS},
\begin{gather*}
|e^{-2\varphi} h_{\rm I}|\le C/t^{(3q+1)/2},
\qquad
|e^{2\varphi} \bar h_{\rm I}|\le C/t^{(3q+1)/2}
\end{gather*}
holds for some $C>0$ on $S_2S_3\cup S_4S_1$.

The estimates of $|e^{-2\varphi} h_{\rm II}|$ on $L^{(1)}$ and of $|e^{2\varphi}\bar h_{\rm II}|$ on $L'^{(1)}$ must be handled
with greater care.
We have \cite[\S~4]{IDNLS}
\begin{gather*}
\psi''(S_j)=(-1)^j 2 S_j^{-2} (2+n/t)^{1/2} (2-n/t)^{1/2}.
\end{gather*}
It can be inf\/initely small and does not lead to a~reasonably good estimate.
We would rather rely on $\psi'''(S_j)$.
The following lemma replaces~\cite[equation~(44)]{IDNLS} in our context.

\begin{lemma}
Let $L^{(1)}(S_j)$ $($resp.\
$L'^{(1)}(S_j))$ be the segment $\subset L^{(1)}$ $($resp.\
$\subset L'^{(1)})$ emanating from~$S_j$.
Let~$d$ be the distance from~$S_j$ to $z\in L^{(1)}(S_j)$ $($resp.\
to $z\in L'^{(1)}(S_j))$.
Then there exists a~positive constant $C'$ such that
\begin{gather}
\label{eq:degree3}
\operatorname{Re} i\psi (z) \ge C' d^3, \quad z\in L^{(1)}(S_j),
\qquad
\operatorname{Re} i\psi (z) \le -C' d^3, \quad z\in L'^{(1)}(S_j).
\end{gather}
\end{lemma}

\begin{proof}
First we assume $j=2$.
In view of Fig.~\ref{fig:phase}, $L^{(1)}$ is in the region $\operatorname{Re} (i \psi)=\operatorname{Re} (t^{-1}\varphi)>0$.
Since $\psi'(S_2)=0$, we have
\begin{gather}
\label{eq:taylor}
i \psi(z)= i \psi(S_2)+\frac{i \psi''(S_2)}{2}(z-S_2)^2+\frac{i \psi'''(S_2)}{6}(z-S_2)^3+\text{higher order terms}
\end{gather}
and $i \psi(S_2)$ is purely imaginary.
It holds that
\begin{gather*}
\psi''(S_2)=2 S_2^{-2} (2+n/t)^{1/2} (2-n/t)^{1/2},
\qquad
\psi'''(S_2) =-12S_2^{-5}-\frac{2n}{it}S_2^{-3}.
\end{gather*}
The segment $L^{(1)}(S_2)$ is tangent to the steepest ascent path of $\varphi=it \psi$, hence also of $i\psi$.
Assume $z\in L^{(1)}(S_2)$.
If $n/t\approx 2$, then $S_2$ is close to $T_1=e^{-\pi i/4}$ and $z-S_2$ is close to $id$.
We have
\begin{gather}
i\psi''(S_2)(z-S_2)^2   \approx 2 (2+n/t)^{1/2} (2-n/t)^{1/2} d^2,
\label{eq:secondorder}
\\
i\psi'''(S_2)(z-S_2)^3   \approx \left(12e^{\pi i/4}+\frac{2n}{t}e^{-3\pi i/4}\right) d^3.
\label{eq:thirdorder}
\end{gather}
The right-hand side of~\eqref{eq:secondorder} is positive.
In estimating $\operatorname{Re} i\psi$ from below, we can neglect the term of degree 2 in~\eqref{eq:taylor}.
On the other hand, the quantity in the parentheses on the right-hand side of~\eqref{eq:thirdorder} has a~positive real
part (close to $4\sqrt{2}$).
Hence we get the f\/irst inequality of~\eqref{eq:degree3}.
By symmetry,~\eqref{eq:degree3} also holds on $L^{(1)}(S_4)$.

In a~similar way, we can show that the second inequality of~\eqref{eq:degree3} holds on $L'^{(1)}(S_2)$.
Notice that $z-S_2\approx d$ on it.
The case $j=2$ is now f\/inished.

By symmetry, we get~\eqref{eq:degree3} for $j=4$.

Since $S_1\approx S_2$, the estimates on $L^{(1)}(S_1)$ and $L'^{(1)}(S_1)$ are similar to those on $L'^{(1)}(S_2)$ and
$L^{(1)}(S_2)$ respectively.
Notice that~$L$ and $L'$ are exchanged.
We get~\eqref{eq:degree3} for $j=1$.
The case $j=3$ follows by symmetry.
\end{proof}

Assume $z\in L^{(1)}(S_j)$.
We have $|\alpha(z)|\le \const d^q$.
By modifying the argument of~\cite[equation~(45)]{IDNLS}, we obtain
\begin{gather}
\label{eq:hIIestimate}
|e^{-2\varphi}h_{\rm II}| \le \const d^q e^{-2C' t d^3} \le \const t^{-q/3} \sup_{\tau>0} \tau^{q/3}e^{-2C' \tau} \le \const t^{-q/3}.
\end{gather}
This kind of estimate obviously holds on any compact subset of $\{\operatorname{Re} \varphi>0\}$.
Hence we get the following lemma.

\begin{lemma}
\begin{gather*}
|e^{-2\varphi}h_{\rm II}| \le \const t^{-q/3}
\qquad
\text{on}
\quad
L^{(1)},
\qquad
|e^{2\varphi}\bar h_{\rm II}| \le \const t^{-q/3}
\qquad
\text{on}
\quad
L'^{(1)}.
\end{gather*}
\end{lemma}

The contribution to $R_n(t)$ by $h_{\rm II}$ and $\bar h_{\rm II}$ on $L^{(1)}\cup L'^{(1)}$, as well as by $h_{\rm I}$ and $\bar
h_{\rm I}$ on $S_3S_2 \cup S_1S_4$, are of order $t^{-l}$ as $t\to\infty$, where $l>0$ is arbitrarily large.
It is justif\/ied by choosing suf\/f\/iciently large~$q$.
We are left with an RHP over $\Sigma^{(2)}=C$, the union of four arcs oriented clockwise.
See Fig.~\ref{fig:Sigma2}.

We follow~\cite[equations~(5.9) and (5.10)]{DZ}.
The new jump matrix $v^{(2)}=v^{(2)}(z)$ is given~by
\begin{gather*}
v^{(2)}=\begin{cases} e^{-\varphi \ad \sigma_3} \left\{
\begin{bmatrix}
1&-\bar r
\\
0 & 1
\end{bmatrix}
\begin{bmatrix}
1 & 0
\\
r & 1
\end{bmatrix}
\right\}
&
\text{on}
\ \
S_1S_2\cup S_3S_4,
\vspace{1mm}\\
  e^{-\varphi \ad \sigma_3} \left\{
\begin{bmatrix}
1 & -R
\\
0 & 1
\end{bmatrix}
\begin{bmatrix}
1 & 0
\\
\bar R & 1
\end{bmatrix}
\right\}
&
\text{on}
\ \
S_2S_3\cup S_4S_1.
\end{cases}
\end{gather*}
Here $S_j S_k$ denotes the minor arc joining $S_j$ and $S_k$.
Let $m^{(2)}$ be the solution to the RHP corresponding to $v^{(2)}$.
Then by~\eqref{eq:reconstruction}, for any $l>0$,
\begin{gather*}
R_n(t)=-\frac{d}{dz}m^{(2)}(z)_{21}\biggl|_{z=0}+O\big(t^{-l}\big).
\end{gather*}
See Section~\ref{sec: Scaling and rotation} for a~more precise (routine) argument based on the Beals--Coifman formula.

\begin{figure}[t]
\centering \includegraphics[width=4.5cm]{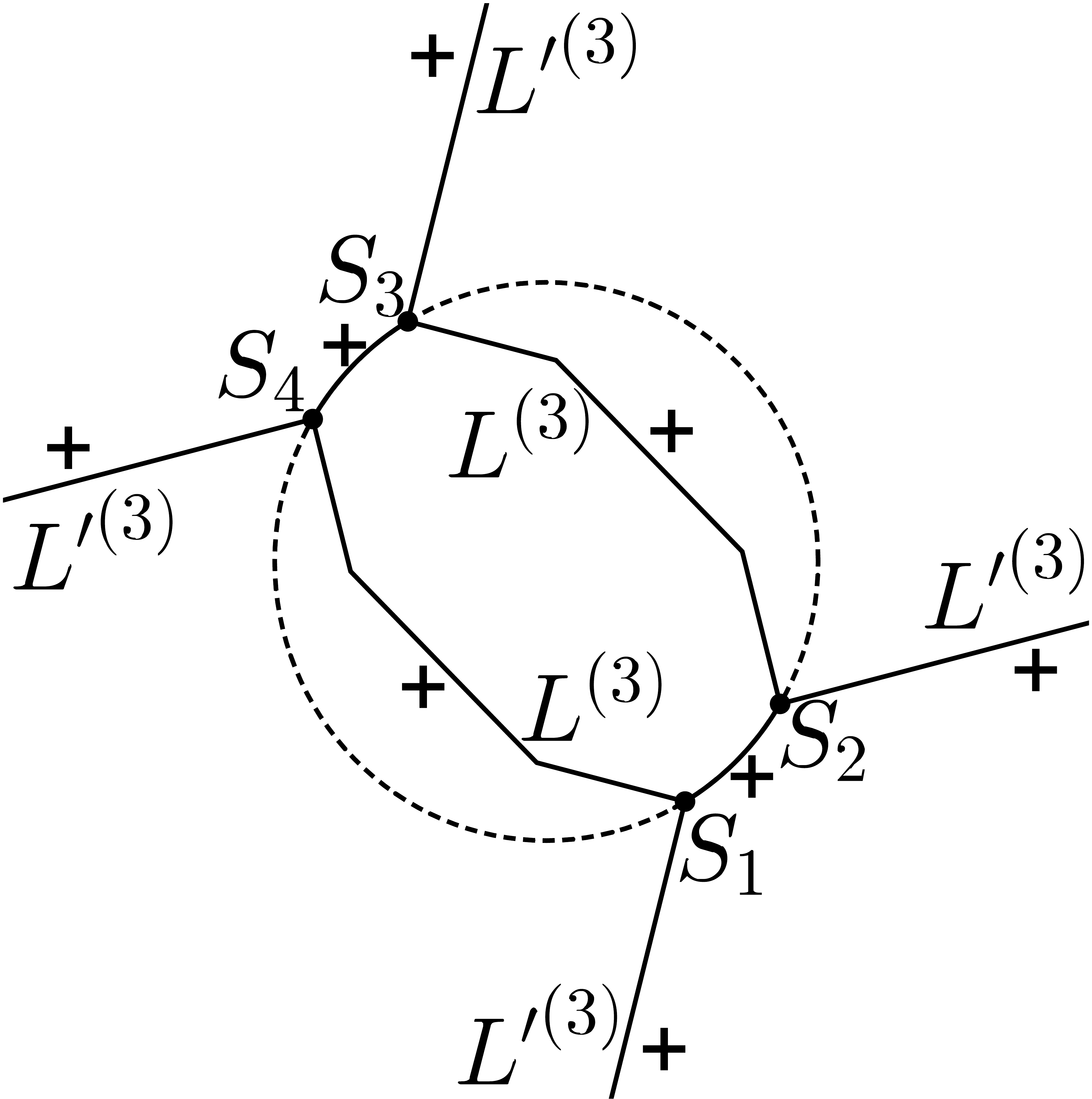} \caption{$\Sigma^{(3)}$.}
\label{fig:Sigma3}
\end{figure}

Let $\Sigma^{(3)}$ be the contour in Fig.~\ref{fig:Sigma3}.
The parts inside and outside the circle are denoted by $L^{(3)}$ and $L'^{(3)}$ respectively.
The latter consists of four half-lines.
Following~\cite[equations~(5.13)--(5.15)]{DZ}, we set
\begin{gather*}
v^{(3)}=v^{(2)}=
\begin{cases}
e^{-\varphi \ad \sigma_3} \left\{
\begin{bmatrix}
1&-\bar r
\\
0 & 1
\end{bmatrix}
\begin{bmatrix}
1 & 0
\\
r & 1
\end{bmatrix}
\right\}
& \text{on}
\ \
S_1S_2\cup S_4S_3,
\vspace{1mm}\\
 e^{-\varphi \ad \sigma_3}
\begin{bmatrix}
1&-R
\\
0&1
\end{bmatrix}
&
\text{on}
\ \
L^{(3)},
\vspace{1mm}\\
 e^{-\varphi \ad \sigma_3}
\begin{bmatrix}
1 & 0
\\
\bar R &1
\end{bmatrix}
&
\text{on}
\ \
L'^{(3)}.
\end{cases}
\end{gather*}
Notice that $v^{(3)}=v^{(3)}(z)\to I$ as $L'^{(3)}\ni z\to \infty$.
The new unknown function $m^{(3)}(z)$ is def\/ined in the usual way: $m^{(3)}(z)\to I$.
It implies that $m^{(3)}(z)=m^{(2)}(z)$ near $z=0$.

\begin{figure}[t]\centering
\begin{minipage}[b]{0.4\hsize} \centering \includegraphics[width=5cm]{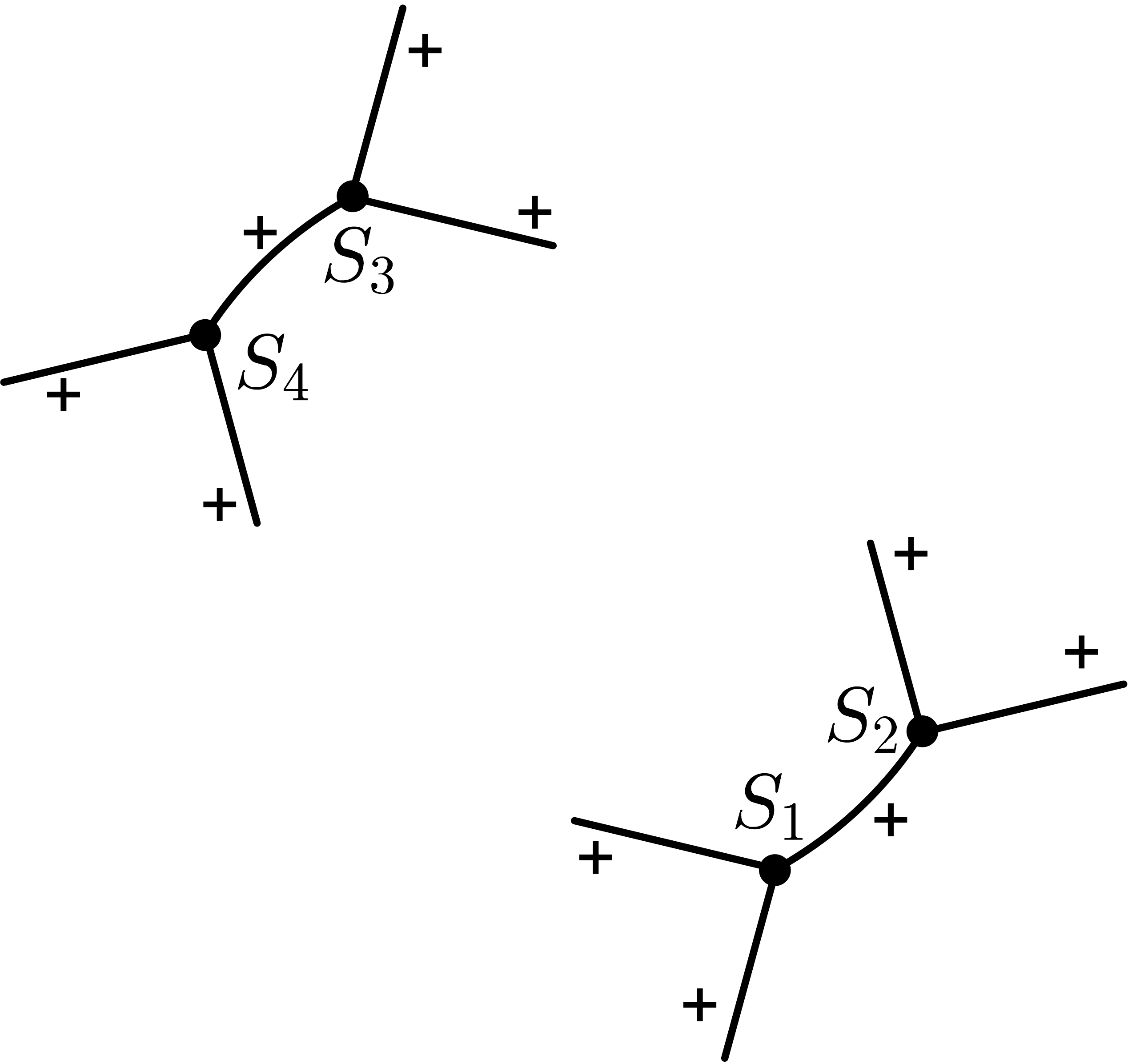} \caption{$\Sigma^{(4)}$.}
\label{fig:Sigma4}
\end{minipage}
\begin{minipage}[b]{0.4\hsize}
\centering \includegraphics[width=5cm]{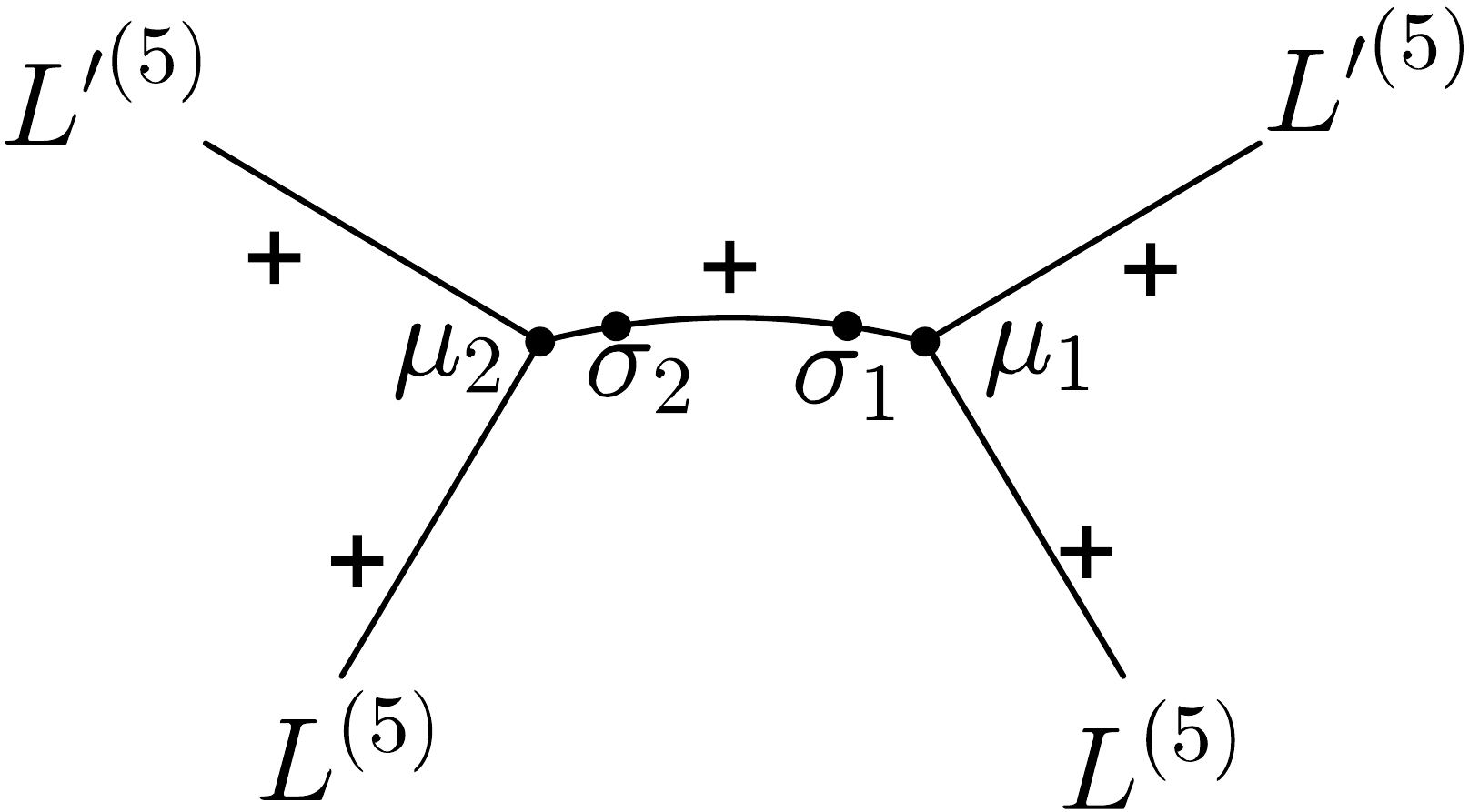}
\caption{$\Sigma^{(5)}$.}
\label{fig:Sigma5}
\end{minipage}
\end{figure}

By using the method of~\cite[\S~7.2, \S~9]{IDNLS} (originally of~\cite[\S~2, \S~3]{DZ}), we can replace $\Sigma^{(3)}$
by the \textit{bounded} contour $\Sigma^{(4)}$ in Fig.~\ref{fig:Sigma4} up to an error of order $O(t^{-1})$, hence
without changing the leading part in the asymptotics.
We can assume that the lengths of the `branches' emanating from the saddle points are independent of~$n$ and~$t$.
The new jump matrix~$v^{(4)}$ equals $v^{(3)}$ on~$\Sigma^{(4)}$ and is the identity matrix elsewhere.

Owing to the technique of~\cite[Proposition~3.66]{DZ} and~\cite[Proposition~9.2]{IDNLS}, the contribution from the two
connected components of $\Sigma^{(4)}$ can be separated out, with an error of $O(t^{-1})$.
Notice that the two terms arising from the two components are actually the same because $r(-z)=-r(z)$ for $z\in C$
(cf.~\cite[Proposition~12.4]{IDNLS}).
It is enough to investigate the lower part (containing $T_1=e^{-\pi i/4}$), which is referred to as
$\Sigma^{(4)}_{\lwr}$.

\section{Scaling and rotation}
\label{sec: Scaling and rotation}

We have
\begin{gather*}
\varphi(T_1) =\frac{i}{4}(-4t+\pi n),
\qquad
\varphi'(T_1)=(2t-n)e^{\pi i/4},
\\
\varphi''(T_1) =i (-2t+n),
\qquad
\varphi'''(T_1)=(-12t+2n)e^{-\pi i/4},
\\
\varphi(z)=\sum\limits_{k=0}^3 \frac{\varphi^{(k)} (T_1)}{k!} (z-T_1)^k +\varphi_4(z),
\qquad
\varphi_4(z)=O\bigl((z-T_1)^4\bigr)
\end{gather*}
near $z=T_1$.
Let $\varepsilon>0$ be such that $\Sigma^{(4)}_{\lwr}$ is within the circle $|z-T_1|=\varepsilon/2$.

Now we def\/ine an operator $\SC$~by
\begin{gather*}
z\mapsto \SC (z)=t^{-1/3} e^{-3\pi i /4} z+T_1,
\qquad
T_1=T_2=e^{-\pi i /4}.
\end{gather*}
Set $\sigma_j=\SC^{-1} (S_j)=t^{1/3} e^{3\pi i/4}(S_j - T_1)$.
We have $S_j - T_1=O\bigl(\sqrt{2-n/t} \bigr)=O(t^{-1/3})$, the latter equality being a~consequence
of~\eqref{eq:Region0308left}.
It follows that $\sigma_j$ is bounded in spite of the magnifying factor~$t^{1/3}$.
There is a~constant $\tilde M>0$ such that $|\operatorname{Re} \sigma_j|<\tilde M$.
Let $\mu_j$ be such that $\operatorname{Re} \mu_j=(-1)^{j-1}\tilde M$ and that $|\SC(\mu_j)|=1$, $\Im\SC(\mu_j)<0$.
We modify $\SC^{-1}\bigl(\Sigma^{(4)}_{\lwr}\bigr)$ without moving the endpoints to get the contour $\Sigma^{(5)}$ in Fig.~\ref{fig:Sigma5}.
The arc $\mu_1 \mu_2$ is a~part of a~circle of radius $t^{1/3}$ and looks like a~segment of length $2\tilde M$ if~$t$ is
large.
The lengths of $L^{(5)}$ and $L'^{(5)}$ are of order $t^{1/3}$ and their directions approach $\pm\pi/4$ or $\pm3\pi/4$
as $t\to\infty$.
We choose $\Sigma^{(5)}$ so that $\SC\bigl(\Sigma^{(5)}\bigr)$ is within the circle $|z-T_1|=\varepsilon$.

We want to approximate $\varphi(\SC(z))$ by a~cubic polynomial which is related to the Painlev\'e II function
(up to a~constant term).
We introduce
\begin{gather*}
\phi=\phi(z)=\frac{i}{4}(-4t+\pi n) +i(-2t+n)t^{-1/3}z+\frac{i(6t-n)t^{-1}}{3}z^3.
\end{gather*}
Then we have $\varphi(\SC(z))=\phi(z)+2^{-1}(2t-n)t^{-2/3}z^2+\varphi_4(\SC(z))$.
The following proposition is an analogue of~\cite[Proposition 10.1]{IDNLS}.

\begin{proposition}
\label{prop:analogueofIDNLS10.1}
Fix a~constant~$\gamma$ with $0<\gamma<1<(6t-n)t^{-1}/3$.
Then on $L^{(5)}$, we have
\begin{gather*}
 \big| e^{-2\varphi(\SC(z))} R(\SC(z))
-e^{-2\phi(z)}\bar r(T_1) \big| \le Ct^{-1/3}
\big|e^{-i\gamma z^3}\big|,
\\
  \big| \SC(z)^{-2} e^{-2\varphi(\SC(z))} R(\SC(z)) -T_1^{-2} e^{-2\phi(z)}\bar r(T_1) \big|
\le Ct^{-1/3} \big|e^{-i\gamma z^3}\big|
\end{gather*}
for some constant $C>0$.
\end{proposition}

\begin{proof}
We show only the latter inequality; the former is easier.
We have
\begin{gather*}
e^{i\gamma z^3} \bigl[\SC(z)^{-2} e^{-2\varphi(\SC(z))} R(\SC(z)) -T_1^{-2}e^{-2\phi(z)}\bar
r(T_1) \bigr]
\\
\qquad{}
= e^{-i\gamma z^3} \bigl[\SC(z)^{-2} ER(\SC(z)) -T_1^{-2}e^{-2\phi(z)+2i\gamma z^3} \bar r(T_1) \bigr],
\end{gather*}
where $E=\exp(-2\varphi(\SC(z))+2i\gamma z^3)$.
Each factor is uniformly bounded.
Notice that $\SC(z)$ remains in the~$\varepsilon$-neighborhood of $T_1$.

Set $f(w)=w^{-2}$.
For any f\/ixed~$z$, $\SC(z)^{-2}=f(\SC(z))$ and $R(\SC(z))$ tend to $T_1^{-2}$ and $\bar r(T_1)$
respectively as $t\to\infty$.
This convergence is uniform on $L^{(5)}$ in the following sense:
\begin{gather*}
 \big| e^{-i\gamma z^3} \bigl[\SC(z)^{-2} -T_1^{-2} \bigr] \big| \le \bigl|e^{-i\gamma z^3}\bigr|
\bigl|t^{-1/3}e^{-3\pi i /4} z\bigr| \sup_{|w-T_1|\le \varepsilon} |f'(w)| \le \const t^{-1/3},
\\
 \big| e^{-i\gamma z^3} \bigl[R(\SC(z)) -\bar r(T_1) \bigr] \big| \le \bigl|e^{-i\gamma z^3}\bigr|
\bigl|t^{-1/3}e^{-3\pi i /4} z\bigr| \sup_{|w-T_1|\le \varepsilon} |R'(w)| \le \const t^{-1/3}.
\end{gather*}
We have used the fact that $e^{-i\gamma z^3}z$ is bounded on either branch of $L^{(5)}$.

Since $e^{-i\gamma z^3}z^j$ ($j=2, 4$) is bounded and $2t-n=O(t^{1/3})$, we have
\begin{gather*}
\big| e^{-i\gamma z^3} \big(E-e^{-2\phi(z) +2i \gamma z^3}\big) \big|
\\
\qquad{}
\le \bigl|e^{-i\gamma z^3}\bigr| \sup_{0\le s\le 1} \left| \frac{d}{ds} \exp \big({-}2\phi(z) +2i \gamma z^3
+s\bigl[(2t-n)t^{-2/3}z^2-2\varphi_4 (\SC(z)) \bigr] \big) \right|
\\
\qquad{}
\le C \bigl|e^{-i\gamma z^3}\bigr| \big[(2t-n)t^{-2/3}|z|^2 +2 |t^{-1/3}z|^4 \big] \le Ct^{-1/3}.
\end{gather*}
Combining the three estimates above, we can derive the desired inequality.
\end{proof}

The factorization problem on $\Sigma^{(4)}_{\lwr}$ is equivalent, up to the change of variables
$z\mapsto\SC(z)$, to one on $\Sigma^{(5)}$, where the jump matrix $v^{(5)}=v^{(5)}(z)$ is
\begin{gather*}
v^{(5)}(z)=
\begin{cases}
e^{-\varphi (\SC(z)) \ad \sigma_3} \left\{
\begin{bmatrix}
1 & -\bar r (\SC(z))
\\
0 &1
\end{bmatrix}
\begin{bmatrix}
1 & 0
\\
r (\SC(z)) &1
\end{bmatrix}
\right\}
&
\text{on}
\ \
\sigma_2 \sigma_1,
\vspace{1mm}\\
 e^{-\varphi (\SC(z)) \ad \sigma_3} \left\{
\begin{bmatrix}
1 & -R (\SC(z))
\\
0&1
\end{bmatrix}
\begin{bmatrix}
1 & 0
\\
\bar R (\SC(z))&1
\end{bmatrix}
\right\}
&
\text{on}
\ \
\sigma_1\mu_1\cup \mu_2\sigma_2,
\vspace{1mm}\\
 e^{-\varphi (\SC(z)) \ad \sigma_3}
\begin{bmatrix}
1 & -R (\SC(z))
\\
0 &1
\end{bmatrix}
&
\text{on}
\  \
L^{(5)},
\vspace{1mm}\\
 e^{-\varphi (\SC(z)) \ad \sigma_3}
\begin{bmatrix}
1 & 0
\\
\bar R (\SC(z)) &1
\end{bmatrix}
&
\text{on}
\ \
L'^{(5)}.
\end{cases}
\end{gather*}
Notice that $v^{(5)}$ is smooth across $\sigma_1$ and $\sigma_2$ to any desired order (choose~$k$ suf\/f\/iciently large).

\begin{figure}[t]\centering
\includegraphics[width=5.0cm]{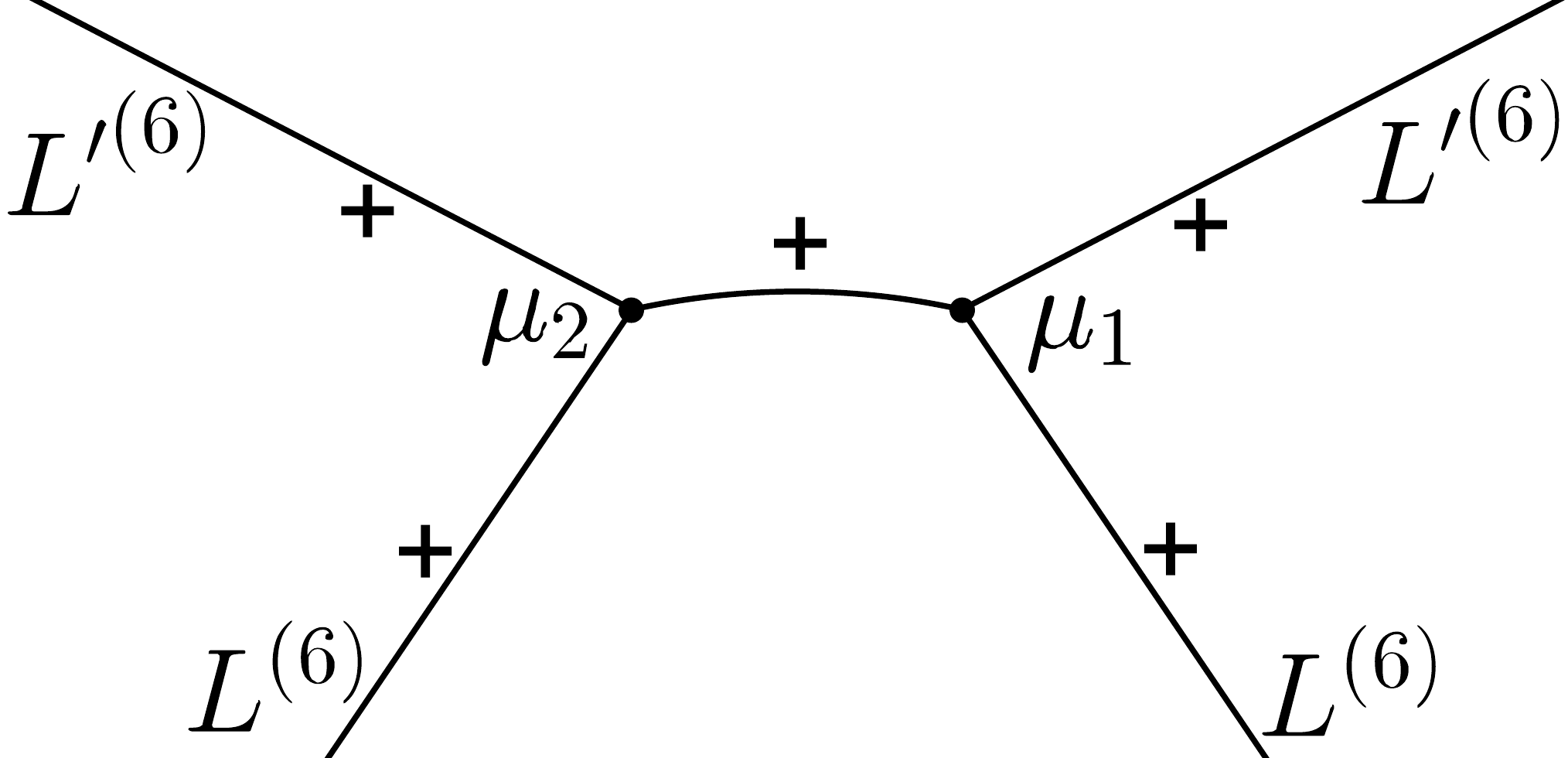} \caption{$\Sigma^{(6)}$.}
\label{fig:Sigma6}
\end{figure}

Let $\Sigma^{(6)}$ be the contour in Fig.~\ref{fig:Sigma6} obtained by extending $L^{(5)}$ and $L'^{(5)}$ inf\/initely.
Then we can regard $v^{(5)}(z)$ as a~jump matrix on $\Sigma^{(6)}$: set $v^{(5)}=I$ on
$\Sigma^{(6)}\setminus\Sigma^{(5)}$.
Because of Proposition~\ref{prop:analogueofIDNLS10.1} (and its variants about $L'^{(5)}$ and $\mu_1 \mu_2$), the jump
matrix $v^{(5)}(z)$ is approximated by $v^{(6)}(z)$ up to an error of order $O(t^{-1/3})$, where
\begin{gather*}
v^{(6)}(z)=
\begin{cases}
e^{-\phi(z) \ad \sigma_3} \left\{
\begin{bmatrix}
1&-\bar r(T_1)
\\
0 & 1
\end{bmatrix}
\begin{bmatrix}
1 & 0
\\
r(T_1) & 1
\end{bmatrix}
\right\}
&
\text{on}
\ \
\mu_1 \mu_2,
\vspace{1mm}\\
 e^{-\phi(z)\ad \sigma_3}
\begin{bmatrix}
1 & -\overline{r}(T_1)
\\
0&1
\end{bmatrix}
&
\text{on}
\ \
L^{(6)},
\vspace{1mm}\\
 e^{-\phi(z) \ad \sigma_3}
\begin{bmatrix}
1&0
\\
r(T_1) &1
\end{bmatrix}
&
\text{on}
\ \
L'^{(6)}.
\end{cases}
\end{gather*}

We rescale by the factor $\alpha=[12t/(6t-n)]^{1/3}>0$, which satisf\/ies $\alpha^3 t^{-1} (6t-n)/3=4$ and tends to
$3^{1/3}$ as $t\to\infty$.
We have
\begin{gather*}
\phi(\alpha z)=\frac{i}{4}(-4t+\pi n) +4i \left\{z^3+\frac{\alpha t^{-1/3}}{4}(-2t+n) z \right\}.
\end{gather*}
Set $p= i (-4t+\pi n)/4, q= \alpha t^{-1/3}(-2t+n)/{4} =2^{-4/3} 3^{1/3} (6t-n)^{-1/3} (-2t+n) $, then we have
\begin{gather*}
\phi(\alpha z)=p+4i(z^3+qz), \qquad p\in i\mathbb{R}.
\end{gather*}
We have normalized the coef\/f\/icient of $z^3$.
The term $4i(z^3+qz)$ will play an important role in Section~\ref{sec:Painleve}.

\begin{figure}[t]
\centering \includegraphics[width=7cm]{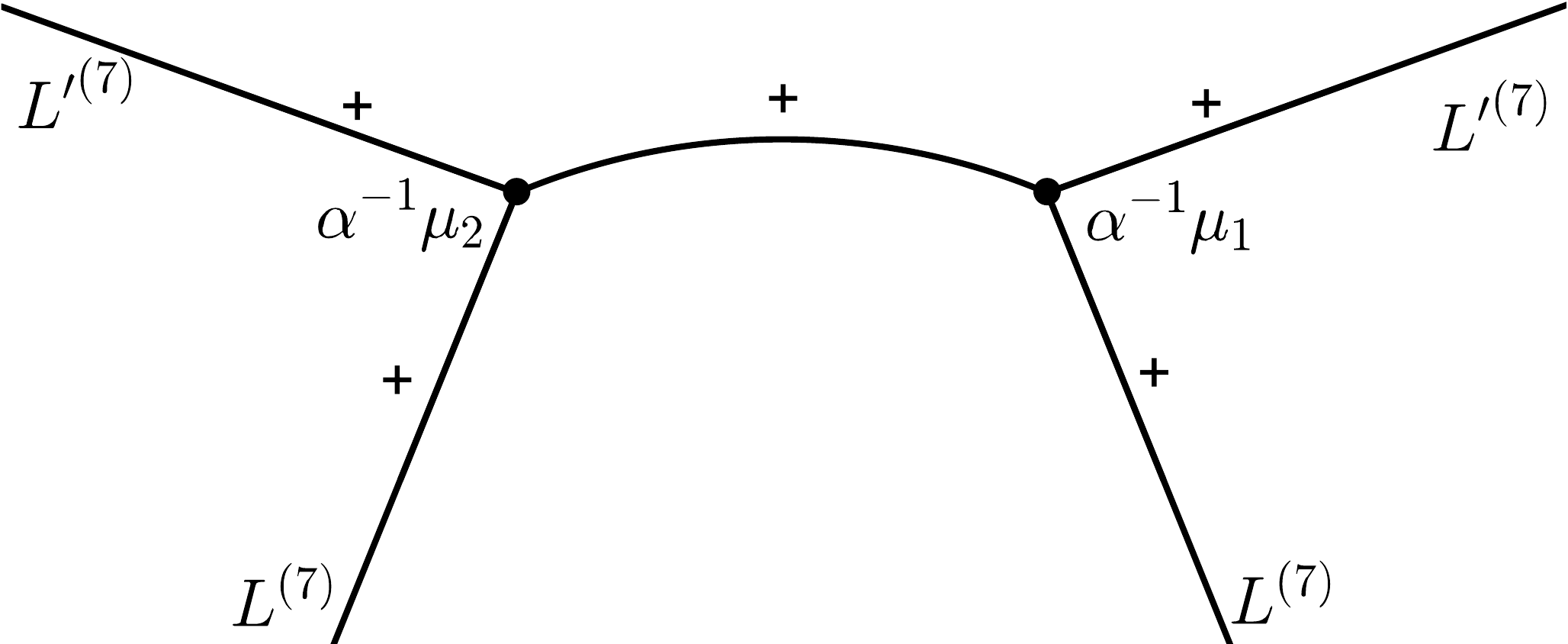} \caption{$\Sigma^{(7)}=\Sigma^{(8)}$.}
\end{figure}

The jump matrix $v^{(7)}(z)=v^{(6)}(\alpha z)$ on $\Sigma^{(7)}=\alpha^{-1}\Sigma^{(6)}$ is given~by
\begin{gather*}
v^{(7)}(z) = \begin{cases} e^{-[p+4i(z^3+qz)]\ad \sigma_3} \left\{
\begin{bmatrix}
1&-\bar r(T_1)
\\
0 & 1
\end{bmatrix}
\begin{bmatrix}
1 & 0
\\
r(T_1) & 1
\end{bmatrix}
\right\}
&
\text{on}
\ \
\big(\alpha^{-1}\mu_1\big)\big(\alpha^{-1}\mu_2\big),
\vspace{1mm}\\
 e^{-[p+4i(z^3+qz)]\ad \sigma_3}
\begin{bmatrix}
1 & -\bar r(T_1)
\\
0& 1
\end{bmatrix}
&
\text{on}
\ \
L^{(7)},
\vspace{1mm}\\
e^{-[p+4i(z^3+qz)]\ad \sigma_3}
\begin{bmatrix}
1 & 0
\\
r(T_1) & 1
\end{bmatrix}
&
\text{on}
\ \
L'^{(7)},
\end{cases}
\end{gather*}
where $(\alpha^{-1}\mu_1)(\alpha^{-1}\mu_2)$ is the arc in $\Sigma^{(7)}$.
We have an RHP $m_+^{(7)}(z)=m_-^{(7)}(z) v^{(7)}(z) $ on $\Sigma^{(7)}$.

We want to remove~$p$ in $v^{(7)}(z)$.
(Notice that~$p$ contributes to the oscillatory factor in Theorem~\ref{th:Region0308}.) Set $m^{(8)}(z)=e^{p\kern.8pt
\ad \sigma_3} m^{(7)}(z)$.
Then $m^{(8)}(z)$ is the solution to
\begin{alignat*}{3}
& m_+^{(8)}(z)=m_-^{(8)}(z) v^{(8)}(z)
\qquad &&
\text{on} \quad
\Sigma^{(8)}=\Sigma^{(7)},&
\\
& m^{(8)}(z) \to I
\qquad &&
\text{as}
\quad
z\to\infty,&
\end{alignat*}
where $v^{(8)}(z)=e^{p  \ad \sigma_3} v^{(7)}(z)$.
We have
\begin{gather*}
v^{(8)}(z) =
\begin{cases} e^{-[4i(z^3+qz)]\ad \sigma_3} \left\{
\begin{bmatrix}
1&-\bar r(T_1)
\\
0 & 1
\end{bmatrix}
\begin{bmatrix}
1 & 0
\\
r(T_1) & 1
\end{bmatrix}
\right\}
&
\text{on}
\ \
\big(\alpha^{-1}\mu_1\big)\big(\alpha^{-1}\mu_2\big),
\vspace{1mm}\\
 e^{-[4i(z^3+qz)]\ad \sigma_3}
\begin{bmatrix}
1 & -\bar r(T_1)
\\
0 & 1
\end{bmatrix}
&
\text{on}
\ \
L^{(8)}=L^{(7)},
\vspace{1mm}\\
 e^{-[4i(z^3+qz)]\ad \sigma_3}
\begin{bmatrix}
1 & 0
\\
r(T_1) & 1
\end{bmatrix}
&
\text{on}
\ \
L'^{(8)}=L'^{(7)}.
\end{cases}
\end{gather*}

We have explained steps of reduction in terms of contours and jump matrices.
It should be supplemented with reconstruction formulas (up to some errors) involving integrals.
Recall that each $v^{(j)}$ has a~factorization of the form $v^{(j)}=(I+w_-^{(j)})(I+w_+^{(j)})$, where the diagonal
components of $w_\pm^{(j)}$ is zero.
We have $(I+w_-^{(j)})^{-1}=I-w_-^{(j)}$.

Let
\begin{gather*}
\big(C_\pm^{(j)}f\big)(z) =\int_{\Sigma^{(j)}} \frac{f(\zeta)}{\zeta-z_\pm} \frac{ d \zeta}{2\pi i} =
\lim_{\genfrac{}{}{0pt}{}{y\to z} {y\in\{\pm\textrm{\scriptsize -side of}\Sigma^{(j)}\kern-.5pt \}}} \int_{\Sigma^{(j)}}
\frac{f(\zeta)}{\zeta-y} \frac{ d \zeta}{2\pi i},
\qquad
z\in\Sigma^{(j)},
\end{gather*}
be the Cauchy operators on $\Sigma^{(j)}$.
Def\/ine $C_{w^{(j)}}\colon L^2(\Sigma^{(j)})\to L^2(\Sigma^{(j)})$~by
\begin{gather*}
C_{w^{(j)}}f=C_+^{(j)}\big(f w_-^{(j)}\big)+C_-^{(j)}\big(f w_+^{(j)}\big)
\end{gather*}
for a~$2\times 2$ matrix-valued function~$f$ (cf.~\cite[\S~2]{DZ},~\cite[\S~7]{IDNLS}).
The Cauchy dif\/ferential form is invariant under an af\/f\/ine change of variables: $z=az'+b$ and $\zeta=a\zeta'+b$ imply
$(\zeta-z)^{-1} d\zeta=(\zeta'-z')^{-1} d\zeta'$.
The operator $C_{w^{(j)}}$ commutes with an af\/f\/ine change of variables in the sense that
\begin{gather*}
(C_{w^{(j)}(z)} f(\bullet)) (az'+b)= (C_{w^{(j)}(az'+b)} f(a\bullet +b)) (z').
\end{gather*}
We have
\begin{gather*}
m^{(j)}(z)=I+ \int_{\Sigma^{(j)}} \frac{\left((1-C_{w^{(j)}})^{-1} I\right)(\zeta)w^{(j)}(\zeta)} {\zeta-z}\frac{d\zeta}{2\pi i},
\end{gather*}
where $w^{(j)}(\zeta)=w_+^{(j)}(\zeta)+w_-^{(j)}(\zeta)$.
By~\eqref{eq:reconstruction}, we obtain
\begin{gather*}
R_n(t)= -\int_{\Sigma^{(1)}} z^{-2} \big[\big((1-C_{w^{(1)}})^{-1} I\big)w^{(1)} \big]_{21} (z) \frac{dz}{2\pi i}.
\end{gather*}
Repeated replacement of contours and integrands leads to (cf.~\cite{IDNLS})
\begin{gather*}
R_n(t)=-\int_{\Sigma^{(4)}} z^{-2} \big[\big((1-C_{w^{(4)}})^{-1} I\big) w^{(4)} \big]_{21} (z) \frac{dz}{2\pi i}+O\big(t^{-1}\big)
\\
\phantom{R_n(t)}
 =-2 \int_{\Sigma^{(4)}_{\lwr}} z^{-2} \big[\big((1-C_{w^{(4)}})^{-1} I\big) w^{(4)} \big]_{21} (z) \frac{dz}{2\pi i}+O\big(t^{-1}\big).
\end{gather*}
By repeated af\/f\/ine changes of variables and Proposition~\ref{prop:analogueofIDNLS10.1}, we get
\begin{gather}
R_n(t)=- \frac{2 e^{-3\pi i/4}}{t^{1/3}} \int_{\Sigma^{(5)}} \SC(z')^{-2} \big[\big((1-C_{w^{(5)}})^{-1}
I\big) w^{(5)} \big]_{21} (z') \frac{dz'}{2\pi i} +O\big(t^{-1}\big)
\nonumber
\\
\phantom{R_n(t)}{}
=- \frac{2 e^{-3\pi i/4}}{t^{1/3}} \int_{\Sigma^{(6)}} T_1^{-2} \big[\big((1-C_{w^{(6)}})^{-1} I\big) w^{(6)}
\big]_{21} (z') \frac{dz'}{2\pi i} +O\big(t^{-2/3}\big)
\nonumber
\\
\phantom{R_n(t)}{}
=-\frac{2 e^{-\pi i/4}\alpha}{t^{1/3}} \int_{\Sigma^{(7)}} \big[\big((1-C_{w^{(7)}})^{-1} I\big) w^{(7)}
\big]_{21} (z) \frac{dz}{2\pi i} +O\big(t^{-2/3}\big),
\label{eq:Rnw7}
\end{gather}
where $\alpha=(12t)^{1/3}(6t-n)^{-1/3}>0$.
We have used the fact that $\SC(z')- T_1=O(t^{-1/3})$ and the second resolvent identity.
See~\cite[Remark~7.4]{IDNLS}.

Let us calculate the integral in~\eqref{eq:Rnw7}.
As $z\to\infty$,
\begin{gather}
z\big[\sigma_3, m^{(j)}(z)\big]_{21}\to 2\left[\int_{\Sigma^{(j)}} \big((1-C_{w^{(j)}})^{-1} I\big) w^{(j)} \right]_{21}
(\zeta) \frac{d\zeta}{2\pi i}.
\label{eq:mjinfinity}
\end{gather}
On the other hand, we have $[\sigma_3, m^{(8)}(z)]= e^{p \ad\sigma_3}[\sigma_3, m^{(7)}(z)] $.
These two formulas imply
\begin{gather}
 \left[\int_{\Sigma^{(8)}} \big((1-C_{w^{(8)}})^{-1} I\big) w^{(8)} \right]_{21} (\zeta) \frac{d\zeta}{2\pi i}
=  \left[e^{p \ad\sigma_3} \int_{\Sigma^{(7)}} \big((1-C_{w^{(7)}})^{-1} I\big) w^{(7)} \right]_{21} (\zeta)\frac{d\zeta}{2\pi i}
\nonumber
\\
\qquad{}
=  e^{-2p} \left[\int_{\Sigma^{(7)}} \big((1-C_{w^{(7)}})^{-1} I\big) w^{(7)} \right]_{21} (\zeta) \frac{d\zeta}{2\pi i}.
\label{eq:w7w8}
\end{gather}
By using~\eqref{eq:Rnw7},~\eqref{eq:mjinfinity} and~\eqref{eq:w7w8}, we obtain
\begin{gather}
R_n(t)= -\frac{2 \alpha e^{2p-\pi i/4}}{t^{1/3}} \int_{\Sigma^{(8)}} \big[\big((1-C_{w^{(8)}})^{-1} I\big) w^{(8)}
\big]_{21} (z) \frac{dz}{2\pi i} +O\big(t^{-2/3}\big)
\nonumber
\\
\phantom{R_n(t)}{}
 =\frac{\alpha e^{2p-\pi i/4}}{t^{1/3}} \lim_{z\to\infty} \bigl\{{-}z\big[\sigma_3, m^{(8)}(z)\big]_{21}\bigr\}+O\big(t^{-2/3}\big).
\label{eq:R_nm^{(8)}}
\end{gather}

\section{Time shift}
If $r(T_j)$ is purely imaginary, it is easy to apply the argument of~\cite[p.~359]{DZ} to our case.
Otherwise, we perform the following reduction.
As is proved in~\cite{APT}, the time evolution of the ref\/lection coef\/f\/icient is given~by
\begin{gather*}
r(T_1, t) =r(T_1)\exp \bigl(it(T_1-\bar T_1)^2 \bigr) =r(T_1)\exp(-2it),
\qquad
r(T_1)=r(T_1, 0).
\end{gather*}
Therefore $r(T_1, t_0)$ is purely imaginary for some $t_0$.
The condition to be satisf\/ied is
\begin{gather*}
\arg r(T_1)-2t_0-\pi /2 \in \pi \mathbb{Z}.
\end{gather*}
Notice that~\eqref{eq:Region0308} is preserved if~$t$ is replaced by $t-t_0$.

\section{Painlev\'e function}
\label{sec:Painleve}
We assume that $r(T_1)$ is purely imaginary.
See the previous section for justif\/ication.

\begin{figure}[t]\centering
\begin{minipage}[t]{0.65\hsize}
\centering \includegraphics[width=7.5cm]{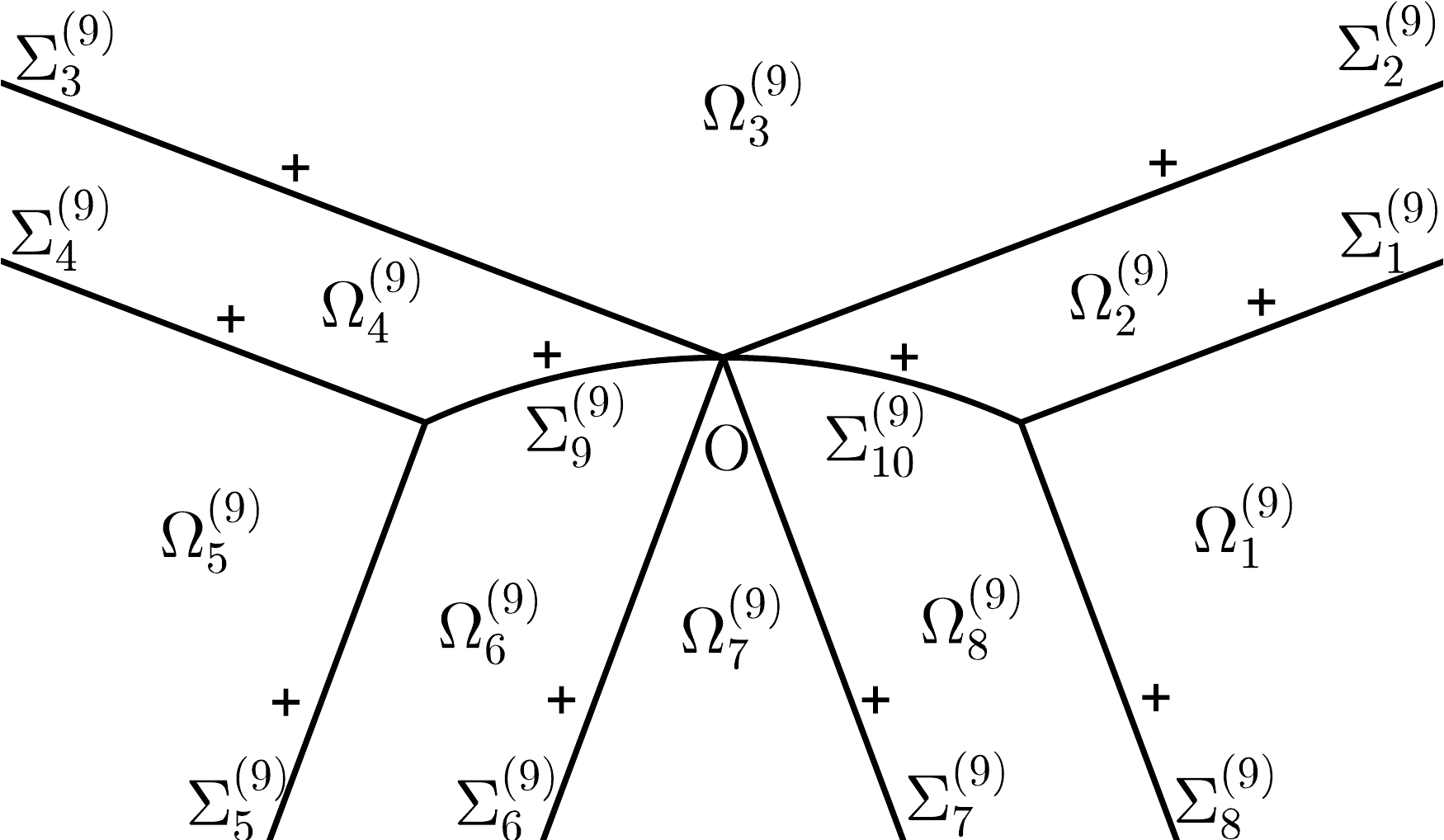} \caption{$\Sigma^{(9)}$.}
\label{fig:Sigma9}
\end{minipage}
\begin{minipage}[t]{0.3\hsize}
\centering \includegraphics[width=2.2cm]{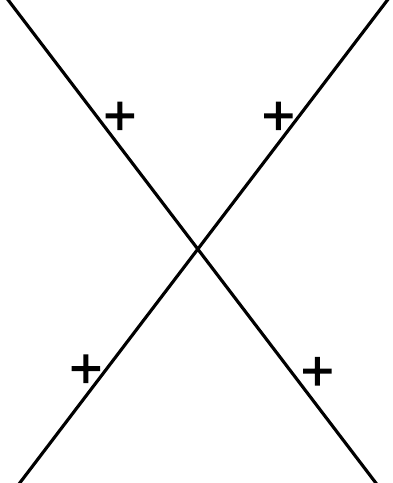} \caption{$\Sigma^{(10)}$.}
\label{fig:Sigma10}
\end{minipage}
\end{figure}

Augment $\Sigma^{(8)}\to \Sigma^{(9)}$ (cf.~\cite[Fig.~5.5]{DZ}) as in Fig.~\ref{fig:Sigma9}.
The contour $\Sigma^{(9)}$ contains four pairs of parallel half-lines.

Def\/ine the new unknown function $m^{(9)}(z)$~by
\begin{gather*}
m^{(9)}(z)=\begin{cases}
m^{(8)}(z),
&
z\in \Omega^{(9)}_1 \cup \Omega^{(9)}_3 \cup \Omega^{(9)}_5 \cup \Omega^{(9)}_7,
\\
m^{(8)}(z) e^{-\{4i(z^3+qz)\}\ad\sigma_3}
\begin{bmatrix}
1&0
\\
-r(T_1)&1
\end{bmatrix}
,
&
z\in \Omega^{(9)}_2 \cup \Omega^{(9)}_4,
\vspace{1mm}\\
 m^{(8)}(z) e^{-\{4i(z^3+qz)\}\ad\sigma_3}
\begin{bmatrix}
1&-\bar r(T_1)
\\
0&1
\end{bmatrix}
,
&
z\in \Omega^{(9)}_6 \cup \Omega^{(9)}_8.
\end{cases}
\end{gather*}
Direct computation shows that $m^{(9)}(z)$ has no jump across $\Sigma^{(9)}_j$ , $j=1, 4, 5, 8, 9, 10$.
Its jump is given by $J_{23}$ across $\Sigma^{(9)}_2 \cup \Sigma^{(9)}_3$ and by $J_{67}$ across $\Sigma^{(9)}_6\cup
\Sigma^{(9)}_7$, where
\begin{gather*}
J_{23} =e^{-\{4i(z^3+qz)\}\ad\sigma_3}
\begin{bmatrix}
1&0
\\
r(T_1)&1
\end{bmatrix}
,
\qquad
J_{67} =e^{-\{4i(z^3+qz)\}\ad\sigma_3}
\begin{bmatrix}
1& -\bar r(T_1)
\\
0&1
\end{bmatrix}
.
\end{gather*}
Thus the RHP is reduced to one along $\Sigma^{(9)}_2\cup \Sigma^{(9)}_3 \cup \Sigma^{(9)}_6 \cup \Sigma^{(9)}_7$.
It is not exactly a~cross, but a~simple deformation enables us to replace it by $\Sigma^{(10)}$ as in
Fig.~\ref{fig:Sigma10}, the counterpart of the contour in~\cite[Fig.~5.6]{DZ}.
In the upper and lower halves, the jump matrix coincides with $J_{23}$ and $J_{67}$ respectively.
We apply the argument in~\cite[pp.~357--360]{DZ}.
In particular, we employ the parameters $\mathrm{p}$, $\mathrm{q}$, $\mathrm{r}$ (roman font) in it.
See Appendix for explanation.
We use $|r(T_j)|<1$ and $r(T_j)+\bar r(T_j)=0$, the latter being true if the time variable~$t$ is replaced by~$t-t_0$
for some~$t_0$ (see the previous section).

We employ the notation explained in Appendix.
We set $\mathrm{p}=r(T_1)$, $\mathrm{q}=-r(T_1)=\overline{r}(T_1)$,
$\mathrm{r}=(\mathrm{p}+\mathrm{q})/(1-\mathrm{p}\mathrm{q})=0$ and consider the solution $u(s; r(T_1), -r(T_1), 0)$ to
the Painlev\'e II equation $u''-su-2u^3=0 $.
Since
\begin{gather*}
4i\big(z^3+qz\big)=\frac{4i}{3}\big(3^{1/3}z\big)^3+i \frac{4q}{3^{1/3}} \big(3^{1/3}z\big),
\end{gather*}
we have
\begin{gather}
\label{eq:m8painleve}
\lim_{z\to\infty} \bigl({-}z \big[\sigma_3, m^{(8)}(z)\big]_{21} \bigr) =\frac{1}{3^{1/3}} u\left(\frac{4q}{3^{1/3}}; r(T_1),
-r(T_1), 0\right).
\end{gather}
We combine~\eqref{eq:m8painleve} with~\eqref{eq:R_nm^{(8)}}.
The result is
\begin{gather*}
R_n(t) =\frac{e^{2p-\pi i/4} \alpha}{(3t)^{1/3}} u\left(\frac{4q}{3^{1/3}}; r(T_1), -r(T_1), 0\right) +O\big(t^{-2/3}\big).
\end{gather*}
Theorem~\ref{th:Region0308} holds at least in the region~\eqref{eq:Region0308left}.

\section{Asymptotics in the remaining part of Region B}
\label{sec:righthalf}

We consider the long-time asymptotics in the region
\begin{gather}
2t\le n<2t+M t^{1/3},
\label{eq:Region0308right}
\end{gather}
where $M'$ is an arbitrary positive constant.
It is the `right-hand half' of the Region B def\/ined by~\eqref{eq:Region0308}.

If $2t=n$, then the function $\varphi(z)$ has no saddle points.
Indeed, $S_j$ and $S_{j+1}$ ($j=1$, $3$) coalesce.
If $2t<n$, then $\varphi(z)$ has four saddle points on the line $\operatorname{Re} z+\mathrm{Im }z=0$.
Set $A=2^{-1}(\sqrt{2+n/t}+\sqrt{-2+n/t})$, $A'=2^{-1}(\sqrt{2+n/t}-\sqrt{-2+n/t})$, then the four saddle points are
$\pm e^{-\pi i/4}A$ and $\pm e^{-\pi i/4}A'$.
Notice that $A>1$, $AA'=1$, $0<A'<1$.

\begin{figure}
[t]\centering
\begin{minipage}
[b]{0.49\hsize} \centering \includegraphics[width=6cm]{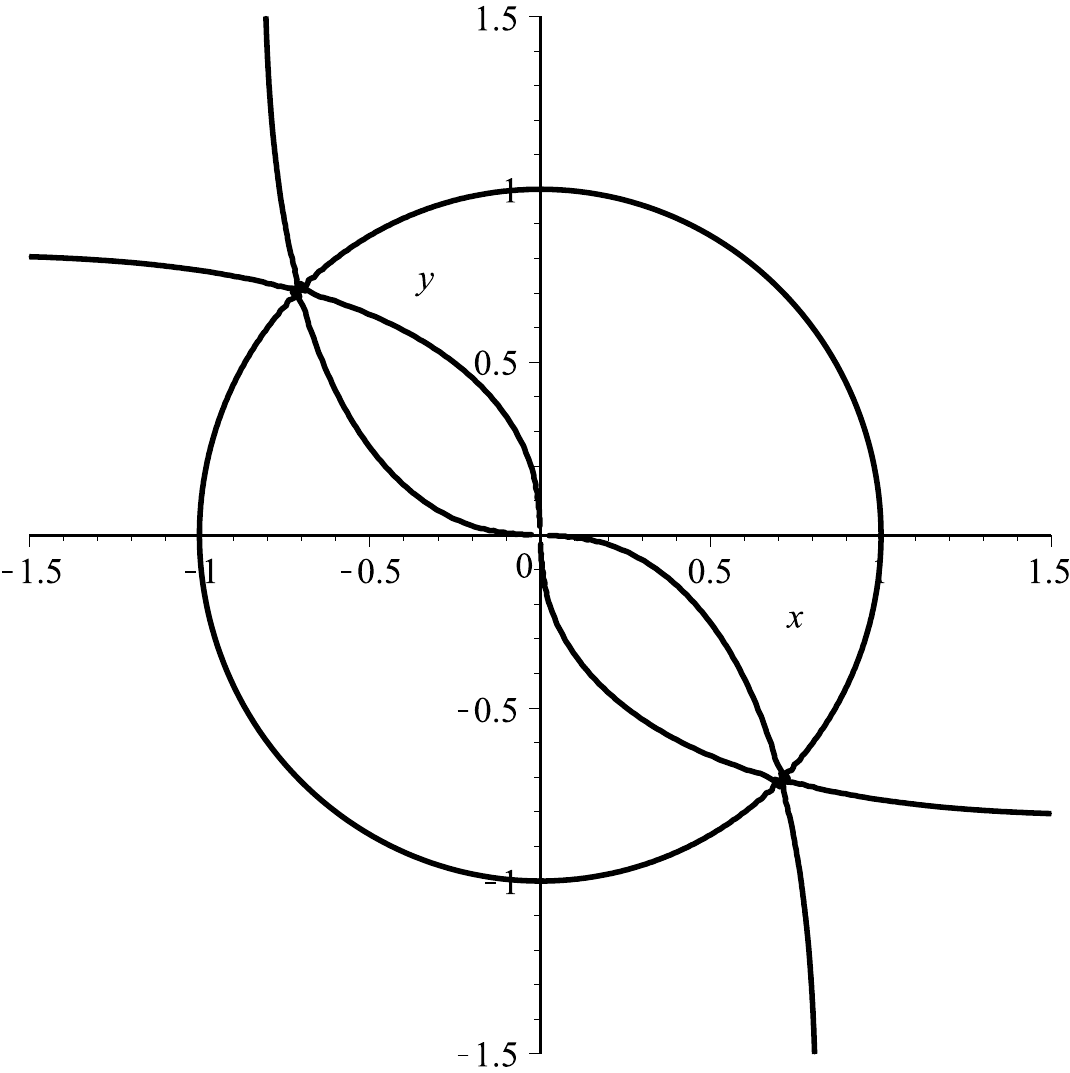} \caption{$n=2t$.}
\label{fig:equal}
\end{minipage}
\begin{minipage}
[b]{0.49\hsize} \centering \includegraphics[width=6cm]{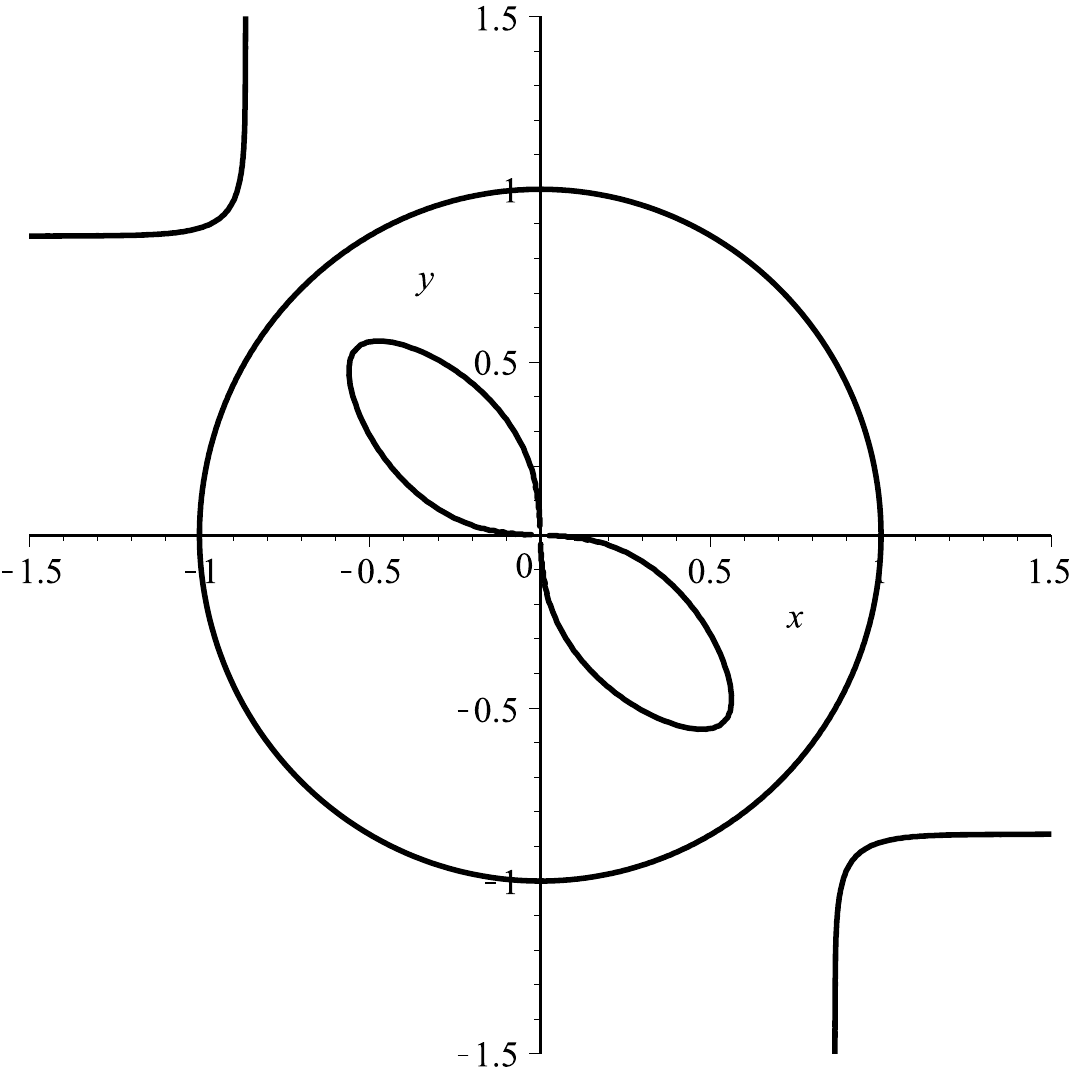} \caption{$n>2t$.}
\label{fig:n-large}
\end{minipage}
\end{figure}

For $z=r e^{i \theta}$ (here~$r$ is not the ref\/lection coef\/f\/icient), we have $\operatorname{Re} \varphi=\frac{-1}{2}t
(r^2-r^{-2}) \sin 2\theta - n \log r.
$ It vanishes for any~$\theta$ if $r=1$.
If $r\ne 1$, the equation $\operatorname{Re} \varphi=0$ is equivalent to saying that
\begin{gather*}
\sin 2 \theta=-\frac{2n}{t}\frac{\log r}{r^2-r^{-2}}.
\end{gather*}
The function ${\log r}/(r^2-r^{-2})$ can be continuously extended to $0<r<\infty$.
It is strictly increasing in $0<r<1$ and is strictly decreasing in $r>1$.
It attains its maximum $1/4$ at $r=1$.
We can calculate the number of solutions~$\theta$ (modulo $2\pi$) for each f\/ixed value of~$r$.
Figs.~\ref{fig:phase},~\ref{fig:equal} and~\ref{fig:n-large} show the curve $\operatorname{Re} \varphi=0$ in the cases
$n<2t$, $n=2t$ and $2t<n$ respectively.

Set $\psi_0=\psi_0 (z)=2^{-1}(z-z^{-1})^2 +2i \log z$.
It is nothing but what~$\psi$ is if $n=2t$.
We employ it as the Fourier variable in the region~\eqref{eq:Region0308right}, not only on the ray $n=2t$.
Then we get a~decomposition like~\eqref{eq:decomp} on $\mathrm{arc}(T_2 T_3)$ and on $\mathrm{arc}(T_4 T_1)$, where
$T_1=T_2=e^{-\pi i/4}$ and $T_3=T_4=e^{3\pi i/4}$.
In the formulas below, $h_{\rm I}$, $h_{\rm II}$ etc.
denote the terms obtained by this decomposition.

\begin{figure}[t]\centering
\begin{minipage}[b]{0.55\hsize} \centering \includegraphics[width=5.7cm]{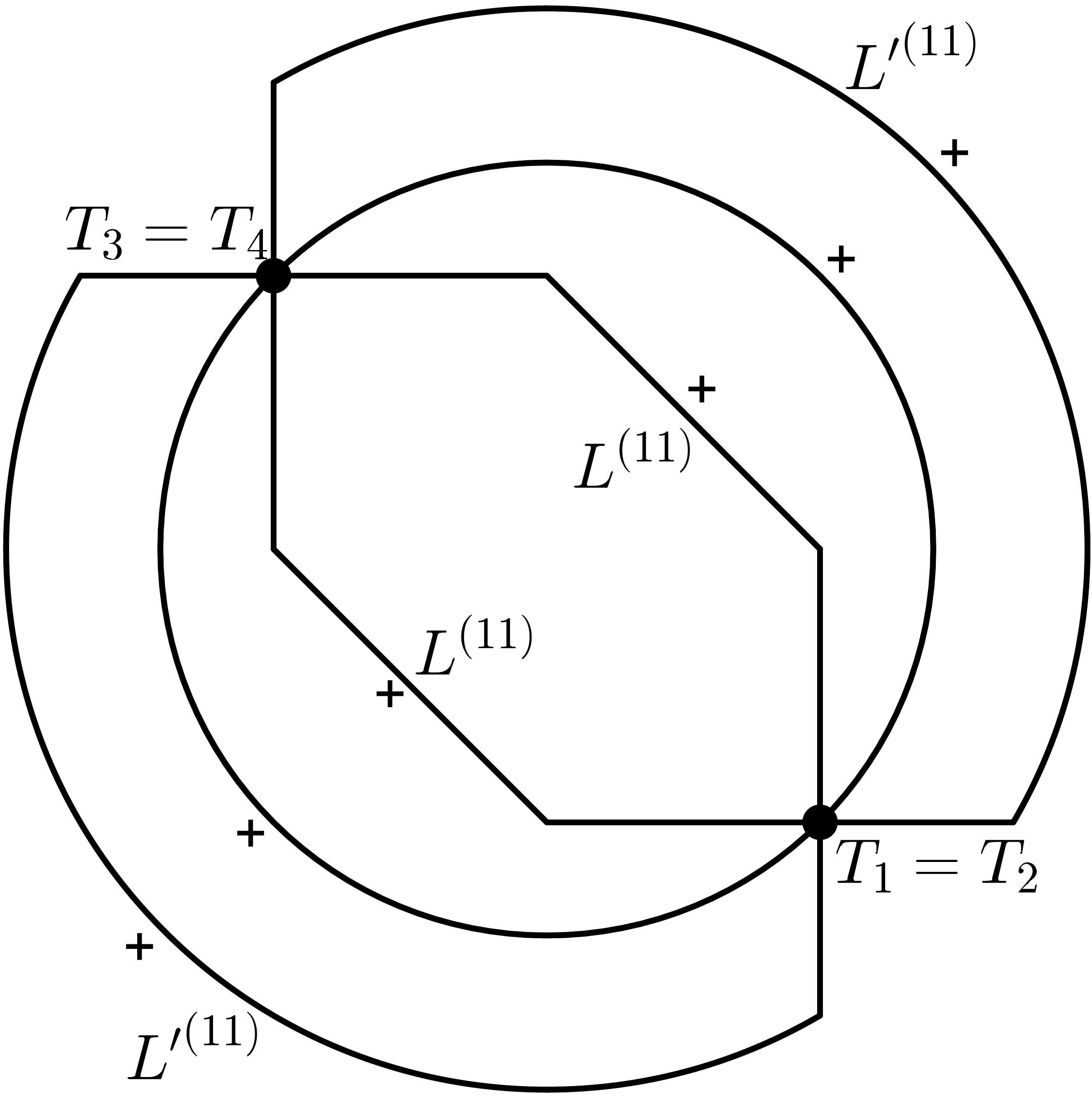} \caption{$\Sigma^{(11)}$.}
\label{fig:Sigma11}
\end{minipage}
\begin{minipage}[b]{0.43\hsize} \centering \includegraphics[width=3.2cm]{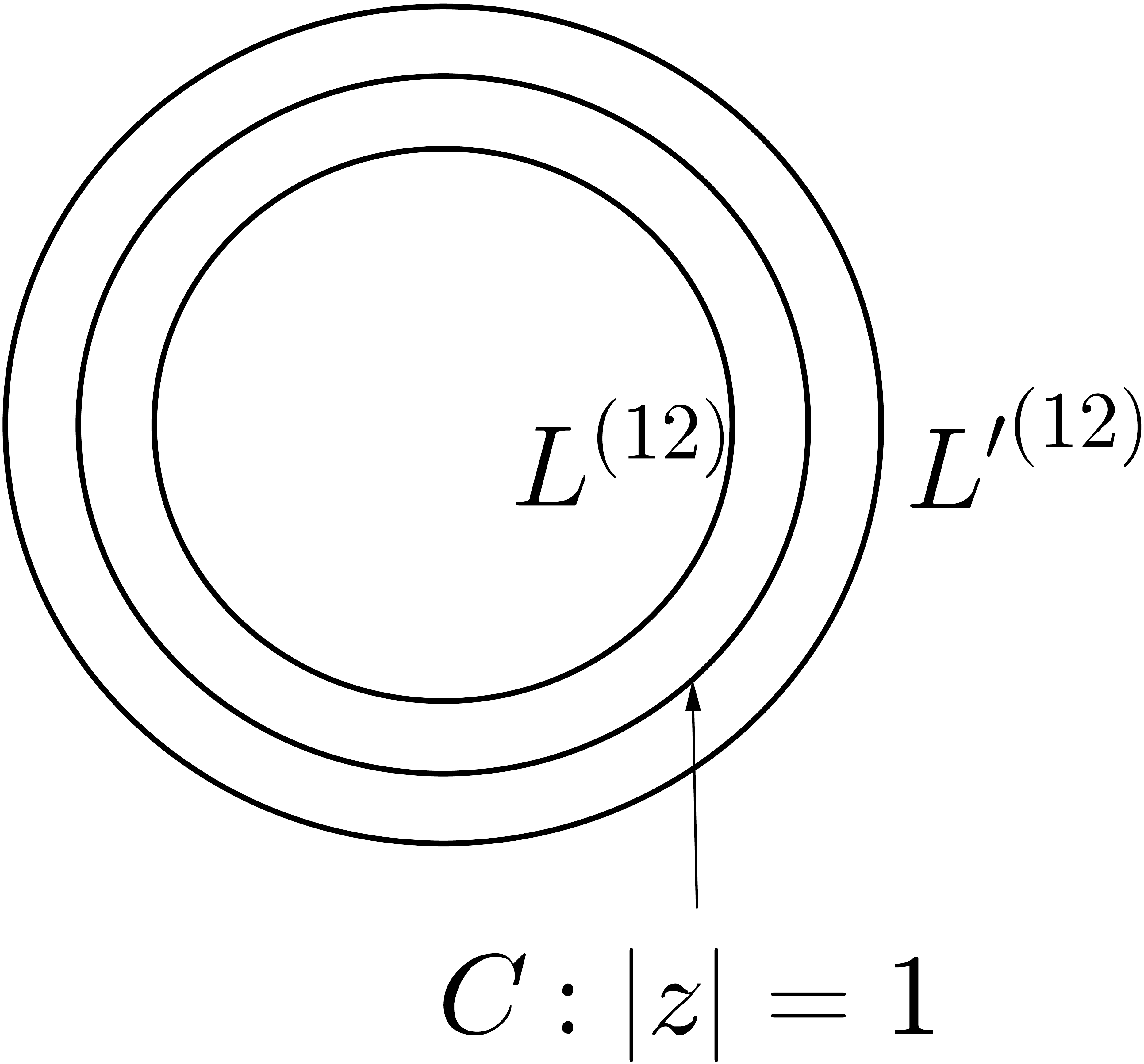} \caption{$\Sigma^{(12)}$.}
\label{fig:Sigma12}
\end{minipage}
\end{figure}

Set $\varphi_0=it\psi_0$.
We have $\operatorname{Re} \varphi>\operatorname{Re} \varphi_0$ in $|z|<1$, and $\operatorname{Re}
\varphi<\operatorname{Re} \varphi_0$ in $|z|>1$.
We introduce a~new contour $\Sigma^{(11)}$ as in Fig.~\ref{fig:Sigma11}.
Notice that $L^{(11)}$ and $L'^{(11)}$ are in $\{\operatorname{Re} \varphi_0>0, |z|<1\}$ and $\{\operatorname{Re}
\varphi_0<0, |z|>1\}$ respectively.

In the same way as~\eqref{eq:hIIestimate}, we can derive estimates of $|e^{-2\varphi_0}h_{\rm II}|$ on $L^{(11)}$.
It is good enough even in the case $n>2t$, because we have $|e^{-2\varphi}h_{\rm II}|\le |e^{-2\varphi_0}h_{\rm II}| $ on
$L^{(11)}$.
By this observation, we can perform a~simplif\/ied version of the argument in the preceding section.
We conclude that Theorem~\ref{th:Region0308} holds in the whole region~\eqref{eq:Region0308}.

\section{Region C}
\label{sec:RegionC}
We consider the case $2t<n\to\infty$.
The four saddle points of~$\varphi$ are not on the circle $C:|z|=1$.
Two of them are inside and the other two are outside.
For $z=re^{i\theta}$, we have
\begin{gather*}
\operatorname{Re} \left[\frac{2\varphi}{n}\right]=-\frac{t}{n}\big(r^2-r^{-2}\big)\sin 2\theta-2\log r.
\end{gather*}
Set $f(r)=n^{-1}t(r^2-r^{-2}) -2\log r$.
If $r>1$, then $\operatorname{Re} [{2\varphi}/{n}]\le f(r)$ and if $r<1$, then $\operatorname{Re} [-{2\varphi}/{n}]\le-f(r)$.
Notice that $f(1)=0$, $f'(1)=2(2t-n)/n<0$.
If $r>1$ is suf\/f\/iciently close to~$1$, then we have $\operatorname{Re} [{2\varphi}/{n}]<0$.
On the other hand, if $r<1$ is suf\/f\/iciently close to~$1$, then we have $\operatorname{Re} [-{2\varphi}/{n}]<0$.

We introduce a~contour as in Fig.~\ref{fig:Sigma12} consisting of three concentric circles $L^{(12)}$, $L'^{(12)}$ and $C:
|z|=1$.
Their radii are suf\/f\/iciently close.
There exists a~positive number $p=p(V_0)<1$ such that $|e^{-2\varphi}|\le p^n$ on $L^{(12)}$ and $|e^{2\varphi}|\le p^n$
on $L'^{(12)}$.

Since $r(z)$ is smooth on $|z|=1$, its complex conjugate can be written in terms of a~Fourier series:
\begin{gather*}
\bar r(z)=\sum\limits_{k=-\infty}^\infty a_k e^{ik\theta} =\sum\limits_{k=-\infty}^\infty a_k z^k.
\end{gather*}
For any $\alpha\in\mathbb{N}$, there exists a~constant $A_\alpha>0$ such that $|a_k|\le A_\alpha/|k|^{\alpha+1}$ holds
for any $k\in\mathbb{Z}$.
If $r(z)$ is analytic, then a~contour deformation leads to $a_k=(2\pi i)^{-1}\int_{|z|=1\pm\varepsilon}z^{-k-1}\bar r(z)
dz$.
So $a_k$ is exponentially decreasing: $|a_k| \le \const (1\pm \varepsilon)^k$.

Set $h_{\rm I} (z)=\sum\limits_{k<-n} a_k z^k$, $h_{\rm II} (z)=\sum\limits_{k\ge -n} a_k z^k$, $\bar h_{\rm I} (z)=\sum\limits_{k<-n}
\bar a_k z^{-k}$ and $\bar h_{\rm II} (z)=\sum\limits_{k\ge -n} \bar a_k z^{-k}$.
We have $r(z)=\bar h_{\rm I} (z)+\bar h_{\rm II}(z)$.
We employ $z^{-k}$ rather than $\bar z^{k}$ with analytic continuation in mind.
Indeed, $h_{\rm II}$ and $\bar h_{\rm II}$ can be analytically continued up to $L^{(12)}$ and $L'^{(12)}$ respectively.
It is easy to see that $h_{\rm I}$ and $\bar h_{\rm I}$ decay faster than any negative power as $n\to \infty$ on the circle, since
they would have fewer terms.
On the other hand, we can show that $e^{-2 \varphi}h_{\rm II}$ and $e^{2 \varphi}\bar h_{\rm II}$ decay exponentially on
$L^{(12)}$ and $L'^{(12)}$ respectively.

We def\/ine a~new jump matrix $v^{(12)}$~by
\begin{gather*}
v^{(12)}= \begin{cases} e^{-\varphi \ad \sigma_3}
\begin{bmatrix}
1 & -h_{\rm II}
\\
0 & 1
\end{bmatrix}
&
\text{on}
\ \
L^{(12)},
\vspace{1mm}\\
  e^{-\varphi \ad \sigma_3} \left\{
\begin{bmatrix}
1 & -h_{\rm I}
\\
0 & 1
\end{bmatrix}
\begin{bmatrix}
1 & 0
\\
\bar h_{\rm I} & 1
\end{bmatrix}
\right\}
&
\text{on}
\ \
|z|=1,
\vspace{1mm}\\
  e^{-\varphi \ad \sigma_3}
\begin{bmatrix}
1 & 0
\\
\bar h_{\rm II} & 1
\end{bmatrix}
&
\text{on}
\ \
L'^{(12)}.
\end{cases}
\end{gather*}
The factorization problem~\eqref{eq:originalRHP1}--\eqref{eq:originalRHP3} is equivalent to the one involving
$v^{(12)}$.
We can show that~$v^{(12)}$ tends to the identity matrix as $n\to \infty$.
The error is smaller than any negative power of~$n$.
Indeed, we have exponential decay on $L^{(12)}$ and $L'^{(12)}$ due to~$\varphi$.
The decay on the circle $|z|=1$ is not so good in general.
If $r(z)$ is analytic, however, $h_{\rm I}$ and $\bar h_{\rm I}$ decay exponentially as $n\to \infty$.
This completes the proof of Theorem~\ref{th:Region0330}.

\appendix

\section{Parametrization of the Painlev\'e functions}
For readers' convenience, we collect some useful facts employed in~\cite{DZ}.

Let $\mathrm{p, q}$ and $\mathrm{r}$ be constants satisfying the constraint $\mathrm{r=p+q+pqr}$.
We def\/ine six matri\-ces~$S_i$~by
\begin{alignat*}{5}
& S_1 =
\begin{bmatrix}
1&0
\\
\mathrm{p} & 1
\end{bmatrix}
,\qquad && S_2 =
\begin{bmatrix}
1& \mathrm{r}
\\
0 & 1
\end{bmatrix}
,\qquad && S_3 =
\begin{bmatrix}
1&0
\\
\mathrm{q} & 1
\end{bmatrix}
,&
\\
& S_4 =
\begin{bmatrix}
1& -\mathrm{p}
\\
0& 1
\end{bmatrix}
,\qquad && S_5 =
\begin{bmatrix}
1&0
\\
-\mathrm{r} & 1
\end{bmatrix}
,\qquad && S_6 =
\begin{bmatrix}
1& -\mathrm{q}
\\
0& 1
\end{bmatrix}
.&
\end{alignat*}

\begin{figure}[t]
\centering \includegraphics[width=4.5cm]{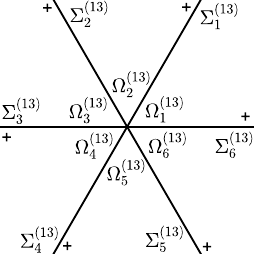} \caption{$\Sigma^{(13)}$.}
\label{fig:Sigma13}
\end{figure}

We introduce the contour $\Sigma^{(13)}$ (the intersection is the origin and all the rays are oriented outward) and the
regions $\Omega^{(13)}_i$ in Fig.~\ref{fig:Sigma13}.
Then we consider the Riemann--Hilbert problem
\begin{gather*}
\Psi_{i+1}(s, z)=\Psi_{i}(s, z) S_j
\qquad
\text{on}
\quad
\Sigma^{(13)}_i (1\le i\le 6),
\end{gather*}
where $\Psi_i$ ($\Psi_7=\Psi_1$) is holomorphic in $\Omega_i^{(13)}$.
It has a~unique solution with the asymptotics
\begin{gather*}
\Psi(s, z)= \left(I+\frac{(\hat Y_i)_1}{z}+\frac{(\hat Y_i)_2}{z^2}+\cdots \right) e^{-([4i/3]z^3+isz)\sigma_3}
\end{gather*}
as $z\to\infty$ in $\Omega^{(13)}_i$.
The function~$u$ def\/ined~by
\begin{gather*}
u=u(s; \mathrm{p, q, r})= - \lim_{z \to \infty} z \bigl[\sigma_3, \hat Y_i (z) \bigr]_{21},
\end{gather*}
where the limit is taken with respect to $z\in \Omega^{(13)}_i$ for any $i\in\{1, \dots, 6\}$, satisf\/ies the Painlev\'e
II equation $u''(s)-su(s)-2u^3(s)=0 $.

\subsection*{Acknowledgments}

This work was partially supported by JSPS KAKENHI Grant Number 26400127.
Parts of this work were done during the author's stay at Wuhan University.
He wishes to thank Xiaofang Zhou for helpful comments and hospitality.

\pdfbookmark[1]{References}{ref}
 \LastPageEnding

\end{document}